%% file: main.tex
\documentclass[sigconf]{acmart}

\usepackage{soul}
\usepackage{algorithm2e}
\usepackage{hyperref}
\usepackage{balance}
\usepackage{tabularx}
\usepackage{float}
\usepackage{multirow}
\usepackage{todonotes}
\usepackage{xltabular}

\definecolor{oxfordblue}{rgb}{0.0, 0.13, 0.28}
\definecolor{harvardcrimson}{rgb}{0.79, 0.0, 0.09}
\definecolor{dartmouthgreen}{rgb}{0.05, 0.5, 0.06}
\definecolor{princetonorange}{rgb}{1.0, 0.56, 0.0}
\definecolor{yaleblue}{rgb}{0.06, 0.3, 0.57}
\definecolor{usccardinal}{rgb}{0.6, 0.0, 0.0}
\definecolor{uclablue}{rgb}{0.33, 0.41, 0.58}
\definecolor{msugreen}{rgb}{0.09, 0.27, 0.23}
\definecolor{cornellred}{rgb}{0.7, 0.11, 0.11}
\definecolor{pomegranate}{RGB}{192, 57, 43}
\definecolor{anti-pomegranate}{RGB}{43,178,192}
\definecolor{alizarin}{RGB}{231, 76, 60} 
\definecolor{anti-belize}{RGB}{185, 41, 56}
\definecolor{belize}{RGB}{41, 128, 185}
\definecolor{peter}{RGB}{52, 152, 219}
\definecolor{green}{RGB}{22, 160, 133}
\definecolor{anti-green}{RGB}{160,22,118}
\definecolor{turquoise}{RGB}{26, 188, 156}
\definecolor{pumpkin}{RGB}{211, 84, 0}
\definecolor{anti-pumpkin}{RGB}{0,22,211}
\definecolor{carrot}{RGB}{230, 126, 34}
\definecolor{wisteria}{RGB}{142, 68, 173}
\definecolor{anti-wisteria}{RGB}{99,173,68}
\definecolor{amethyst}{RGB}{155, 89, 182}
\definecolor{nephritis}{RGB}{39, 174, 96}
\definecolor{anti-nephritis}{RGB}{174,39,117}

\SetKwInput{KwInput}{Input}
\SetKwInput{KwOutput}{Output}
\SetKwInput{KwName}{Name}

\usepackage{mathtools}

\AtBeginDocument{%
  \providecommand\BibTeX{{%
    \normalfont B\kern-0.5em{\scshape i\kern-0.25em b}\kern-0.8em\TeX}}}

\copyrightyear{2024}
\acmYear{2024}

\begin{document}


\title[Flowy: Supporting UX Design Decisions with AI-Annotated Multi-Screen User Flows]{Flowy: Supporting UX Design Decisions Through AI-Driven Pattern Annotation in Multi-Screen User Flows}


\author{Yuwen Lu}
\affiliation{%
  \institution{University of Notre Dame}
  \city{Notre Dame}
  \state{IN}
  \country{USA}}
\email{ylu23@nd.edu}

\author{Ziang Tong}
\affiliation{%
  \institution{University of Notre Dame}
  \city{Notre Dame}
  \state{IN}
  \country{USA}}
\email{ztong2@nd.edu }

\author{Qinyi Zhao}
\affiliation{%
  \institution{University of Washington}
  \city{Seattle}
  \state{WA}
  \country{USA}}
\email{qyzhao@uw.edu}

\author{Yewon Oh}
\affiliation{%
  \institution{University of Notre Dame}
  \city{Notre Dame}
  \state{IN}
  \country{USA}}
\email{yoh2@nd.edu}

\author{Bryan Wang}
\affiliation{%
  \institution{University of Toronto}
  \city{Toronto}
  \state{ON}
  \country{Canada}}
\email{bryanw@dgp.toronto.edu}

\author{Toby Jia-Jun Li}
\affiliation{%
  \institution{University of Notre Dame}
  \city{Notre Dame}
  \state{IN}
  \country{USA}}
\email{toby.j.li@nd.edu}

\begin{teaserfigure}\includegraphics[trim=0cm 0cm 0cm 0cm, clip=true, width=\textwidth]{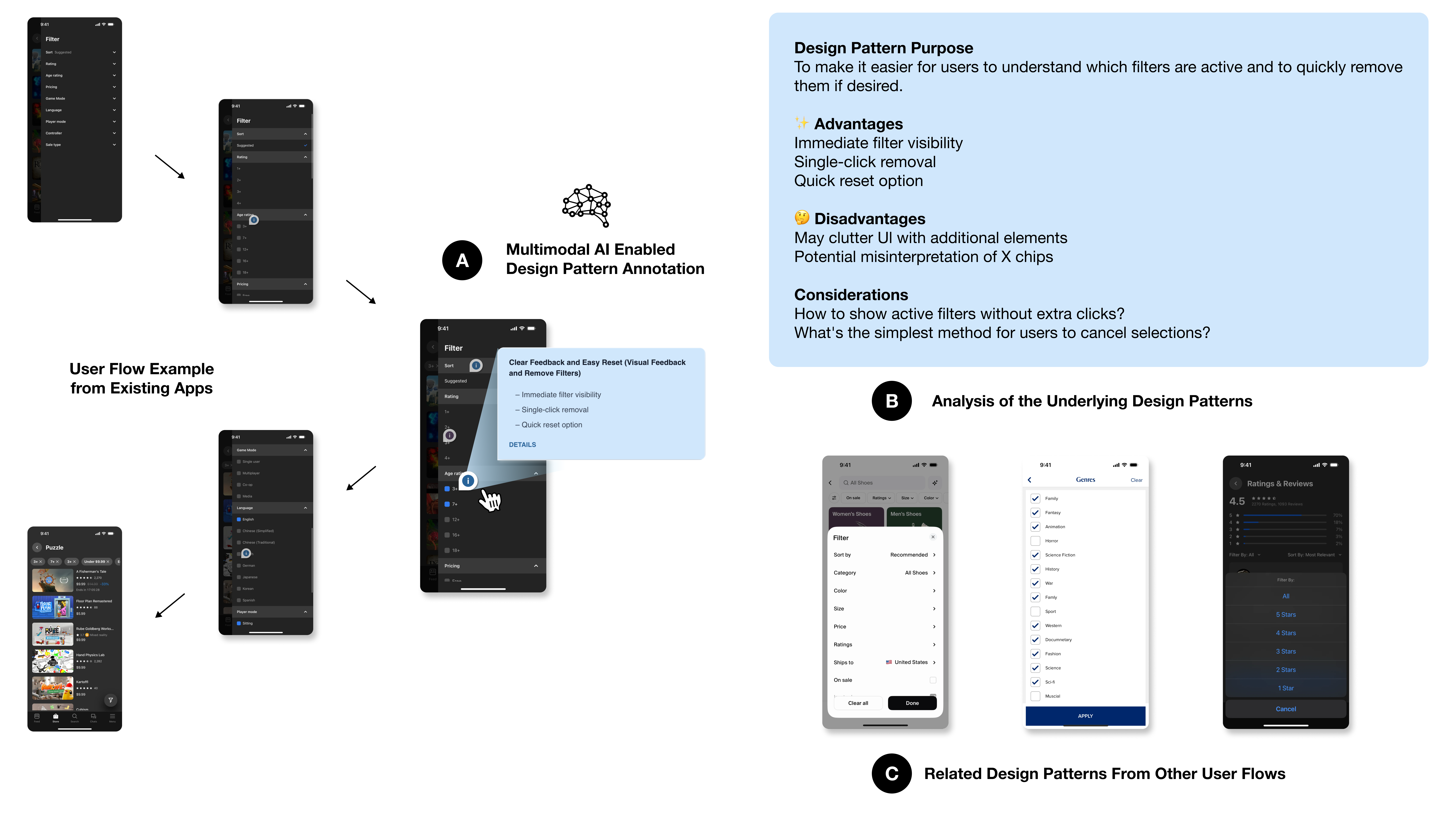}
    \caption{Overview of the Flowy system's approach of UX design support. Flowy system uses a pipeline enabled by multimodal AI to annotate design patterns in high-quality user flow examples from existing apps (A). It provides designers with an analysis of the underlying design patterns in the examples (B) and surfaces other high-quality design examples with related patterns (C). This scaffolds UX designers' decision making processes and accelerates their information foraging process in design ideation, facilitating a holistic understanding of the design space.}
    \label{fig:teaser_figure}
    \Description{The figure consists of multiple smartphone screenshots and user interface elements arranged in a collage layout. The central part of the figure (labeled A) shows the system's multimodal AI-enabled design pattern annotation feature. It displays a smartphone screenshot with various UI components, such as buttons and text fields, along with a panel on the right side showing the purpose, advantages, disadvantages, and considerations of a specific design pattern. The left side of the figure showcases user flow examples from existing apps on smartphone screens. The right side of the figure (labeled B) presents the analysis of the underlying design patterns. It includes three smartphone screenshots displaying different design patterns, such as a filter screen, a cart screen, and a bottom sheet with buttons and sliders. Arrows connect the user flow examples to the central AI annotation feature and then to the design pattern analysis screens, illustrating the system's workflow in supporting designers' UX inspiration foraging and informed decision-making process.}
\end{teaserfigure}

\begin{abstract}

Many recent AI-powered UX design tools focus on generating individual static UI screens from natural language. However, they overlook the crucial aspect of interactions and user experiences across multiple screens. Through formative studies with UX professionals, we identified limitations of these tools in supporting realistic UX design workflows. In response, we designed and developed Flowy, an app that augments designers' information foraging process in ideation by supplementing specific user flow examples with distilled design pattern knowledge. Flowy utilizes large multimodal AI models and a high-quality user flow dataset to help designers identify and understand relevant abstract design patterns in the design space for multi-screen user flows. Our user study with professional UX designers demonstrates how Flowy supports realistic UX tasks. Our design considerations in Flowy, such as representations with appropriate levels of abstraction and assisted navigation through the solution space, are generalizable to other creative tasks and embody a human-centered, intelligence augmentation approach to using AI in UX design.

\end{abstract}





\settopmatter{printfolios=true}

\maketitle

\input{sections/1-introduction}
\input{sections/2-related-work}
\input{sections/3-formative-study}
\input{sections/4-design-goals}
\input{sections/5-system-design}
\input{sections/6-user-study}
\input{sections/7-discussion}
\input{sections/8-conclusion-ack}

\balance
\bibliographystyle{ACM-Reference-Format}
\bibliography{references}

\clearpage
\onecolumn
\appendix

\section{Participant Demographics}
\subsection{Formative Study}
\begin{table*}[!htbp]  
\small
\begin{tabularx}{\textwidth}{XXXlXX}
\toprule
\textbf{Gender} & \textbf{Education} & \textbf{Industry} & \textbf{Job Title} & \textbf{Years in Industry} & \textbf{Company Size} \\
\midrule
Female & Master's degree & Software & Product Designer & 1 to 3 years & 1,000 to 10,000 \\
Female & Master's degree & Software & Senior Product Designer & 3 to 5 years & 100 to 1,000 \\
N/A & Master's degree & Software & UX Designer & 1 to 3 years & More than 10,000 \\
Female & Master's degree & Finance and Insurance & Senior UX Designer & Less than 1 year & More than 10,000 \\
N/A & Master's degree & Software & Senior Product Designer & 3 to 5 years & 10 to 100 \\
\bottomrule
\end{tabularx}
\caption{The demographics of our formative study participants. We used ``Software'' in short for ``Software, Information Services and Data Processing'' for the ease of display.}
\label{tab:formative-demo}
\end{table*}

\subsection{Flowy User Study}

\begin{table*}[!htbp]
\small
\begin{tabularx}{\textwidth}{lXXXXXXl}
\toprule
\textbf{Gender} & \textbf{Education} & \textbf{Industry} & \textbf{Job Title} & \textbf{Industry Experience} & \textbf{Company Size} & \textbf{UX Experience} & \textbf{Used Mobbin} \\
\midrule
N/A & Master's degree & Software & Product designer & Less than 1 year & 100 to 1,000 & Less than 1 year & Yes \\
N/A & Bachelor's degree & Unemployed & interaction design student & Less than 1 year & 1 to 10 & Less than 1 year & Yes \\
N/A & Master's degree & design & product designer & 1 to 3 years & 1,000 to 10,000 & 1 to 3 years & Yes \\
Female & Master's degree & Finance and Insurance & Senior UX designer & 1 to 3 years & More than 10,000 & 1 to 3 years & Maybe \\
Male & Master's degree & Software & Associate UX Designer & 3 to 5 years & 1,000 to 10,000 & 3 to 5 years & Yes \\
Female & Master's degree & Software & Product designer & 3 to 5 years & More than 10,000 & 3 to 5 years & Yes \\
Female & Bachelor's degree & Unemployed & Product Designer & 1 to 3 years & 1 to 10 & 1 to 3 years & Maybe \\
N/A & Master's degree & Scientific or Technical Services & UX Researcher & Less than 1 year & 1 to 10 & Less than 1 year & No \\
\bottomrule
\end{tabularx}
\caption{Demographics of our user study's participants, including gender, educational background, industry, job title, years of experience in the industry, company size, years of experience as a UX designer, and whether they have used Mobbin.com.}
\label{tab:user-study-demo}
\end{table*}



\end{document}

%% file: sections/1-introduction.tex
\section{Introduction}

Recent advances in artificial intelligence (AI) have spurred the development of research prototypes and commercial tools designed to aid in User Interface (UI) and User Experience (UX) design. A prominent group of these UI/UX design support tools automatically generate mid-fidelity UI wireframes~\cite{lu2023ui, feng_layoutgpt_2023, huang_creating_2021} or high-fidelity UI mockups\footnote{A non-exclusive list of non-academic tools include \hyperlink{https://uizard.io/}{Uizard}, \hyperlink{https://www.usegalileo.ai/}{Galileo}, \hyperlink{https://v0.dev/}{V0 by Vercel}, \hyperlink{https://www.magicpatterns.com/}{Magic Patterns}.}~\cite{zhao_guigan_2021, wei2023boosting, mozaffari_ganspiration_2022}. Most of these tools follow a ``prompt-to-UI'' interaction paradigm, similar to the prompt-based image generation paradigm in diffusion models~\cite{dhariwal2021diffusion, ho2020denoising}, where a user provides the description of a UI and the system returns a high-fidelity mockup, aiming to reduce designers' manual efforts in prototyping.

However, existing prompt-to-UI tools predominantly focus on \textit{individual} static UI design, thereby not fully addressing the broader and crucial aspect of \textit{experience design} and \textit{interactions} in UX \cite{norman1998definition, norman2013design}, which is oftern embodied in user flows that extend beyond the scope of individual UI screens. This observation led us to hypothesize that the current prompt-to-UI tools might not adequately support the workflows of professional UX designers.

To validate this hypothesis, we conducted formative studies with professional UX designers to understand the limitations of existing prompt-to-UI tools in real-world UX design workflows. We identified significant gaps between these tools' capabilities and the needs of designers, including: (1) the inability to generate designs based on corporate design systems~\cite{frost_atomic_2016, churchill2019scaling}, (2) the lack of design rationale or justifications, and (3) the narrow focus on individual UI screens, overlooking critical UX details and contexts. We also learned about the important role of \textit{design patterns} in helping UX designers understand the design considerations and potential design space for different product features. Furthermore, we discovered that the UX design process, particularly the ideation phase, resembles an \textit{information foraging process}, where individuals use strategies to maximize the rate of acquiring valuable information while minimizing search costs~\cite{pirolli1999information}. For example, UX designers often need to explore a vast design space of user flows and interactions, implicitly evaluating and selecting \textit{examples} most relevant to their design context. Motivated by these findings, we aim to develop tools that \textbf{enhance the designers' inspiration foraging process during ideation}, rather than building fully automatic UI generation tools, which have fallen short in supporting the UX design process.

We present Flowy, an interactive system that enhances and streamlines UX ideation through the automated analysis of UX design patterns. Flowy automatically identifies three types of patterns: component layout, user interaction, and product feature-specific patterns, aiding designers in decision-making across diverse scenarios. To accomplish this, Flowy leverages a novel pipeline based on multimodal generative AI models, including vision-language models, and utilizes a high-quality commercial dataset, featuring comprehensive user flow exmaples of mobile UIs from Mobbin\footnote{https://mobbin.com/}. With Flowy, designers can extend their vision beyond the design of individual UI screen by accessing information on design patterns across multiple screens in \textit{user flows}.








We conducted a user study with eight professional UX designers to understand the effectiveness of Flowy in supporting realistic UX design tasks. The study results are promising, with participants reporting that Flowy's design pattern analysis and visual comparison of similar examples help guide their attention and facilitate the learning of design patterns. Moreover, through the study, we identified important areas for future work in assisting real-world UI/UX design, such as further integrating product goals or business objectives and prioritizing visuals over textual descriptions in similar tools. Based on these findings, we derive three key design implications that shed light on future research in UX design ideation support tools. More broadly, Flowy exemplifies a intelligence-augmentation approach to human-AI collaboration, with the goal of supporting the cognitive processes of target users to improve task efficiency and effectiveness.


Altogether, Flow provides the following contributions:
\begin{enumerate}
    \item The design and implementation of Flowy, a novel system that supports UX design ideation as an information foraging process;
    \item A technical pipeline leveraging state-of-the-art multimodal AI to automatically annotate design pattern information on user flow design examples;
    \item Study results demonstrating the effectiveness of Flowy, along with three design implications for UX ideation support tools proposed based on the study findings.
\end{enumerate}

%% file: sections/2-related-work.tex
\section{Background and Related Work}

\subsection{Data-Driven UI/UX Design Tools}
There have been long-lasting efforts to build tools that support the UI design process. For example, non-AI tools have attempted to help designers with automatic exploration of design alternatives (e.g. Scout~\cite{swearngin_scout_2020}, SpaceWalker~\cite{zhong2021spacewalker}) and adapting screenshot examples into designs (e.g. Rewire~\cite{swearngin_rewire_2018}). Since UI design is inherently diverse and challenging to capture with rule-based methods, more recently, studies have leveraged data-driven approaches to learn design semantics or patterns from large-scale UI datasets. For instance, the Rico \cite{deka_rico_2017} dataset is one of the most prominent open-source UI datasets for Android. ERICA \cite{deka_erica_2016} offers a collection of user interaction data for mobile UIs, captured during app usage. WebUI \cite{wu_webui_2023} is a dataset for enhancing visual UI understanding with web semantics. Widget Captioning \cite{li_widget_2020} provides language descriptions of individual UI elements, while Screen2Words \cite{wang_screen2words_2021} delivers summaries of entire UI screens. Building on these datasets, prior research has explored machine learning approaches to model UI semantics, achieve computational understanding of UIs, and enable novel interactions with UIs. For example, Screen2Vec \cite{li_screen2vec_2021} employs a self-supervised technique for generating representations of GUI screens in embedding vectors. More recently, there has been significant interest in modeling UIs using foundation models. For instance, Wang et al. \cite{wang_enabling_2023} utilized a large language model to process mobile UIs, enabling diverse conversational interactions with the UI. Conversely, Spotlight \cite{li_spotlight_2023} proposed a vision-only approach for understanding mobile UIs. In this study, Flowy built upon this prior work and proposed a computational pipeline that integrates state-of-the-art multimodal AI models to articulate design pattern information on UX prototypes and provide support for UX designers.

It is noteworthy that the prevalent adoption of design systems in the UX industry has significantly re-shaped the landscape of UX design research. Design systems are centralized collections of pre-defined UI components and design guidelines that ensure design consistency across products and teams~\cite{frost_atomic_2016, churchill2019scaling}. Examples of established design systems in the industry nowadays include Apple's Human Interface Guidelines\footnote{https://developer.apple.com/design/human-interface-guidelines} and Google's Material Design\footnote{https://material.io/}. As a result of the wide adoption of design systems, UX designers focus less on crafting pixel-perfect visual details. Instead, they spend more time creating holistic user experiences that support user goals and product objectives using pre-built UI components~\cite{lu_bridging_2022}. This change in UX practice is often underestimated in academia.
    
\subsection{Ideation with Generative AI}
Generative AI has been utilized to enhance user creativity by quickly providing a wide array of ideas in various formats, including text and visuals, aiding users in overcoming creative blocks and explore new concepts. The ideation process may involve the generation of text that directly articulates ideas \cite{di2022idea, yin2024jamplate, gonzalez2024collaborative, girotra2023ideas, shaer2024ai}. Moreover, ideation can also be based on examples, such as when users generate various images using a text-to-image generation model to explore design examples they prefer \cite{paananen2023using, lamac2023text, oppenlaender2022creativity, ko2023large, brade2023promptify, qu2023sketchdreamer}. While straightforward prompting has been widely used to harness generative AI for idea generation, recent research has explored more systematic approaches to further enhance the ideation process, such as incorporating object-oriented principles into LLM prompting \cite{kim2023cells} and enabling structured exploration and expansion of the design space by leveraging LLM outputs \cite{suh2023structured}. In this work, Flowy takes a less common approach to inspiration. Instead of generating example results to help designers find new inspiration, Flowy focuses on augmenting designers' natural information foraging process for UX inspirations through design pattern analysis.

\subsection{Supporting the Information Foraging Process}
Since its proposal in 2016, the information foraging theory has become a seminal in understanding how humans seek and gather information~\cite{pirolli1999information}. Information foraging theory draws an analogy between how humans search for information and how animals forage for food in the wild. It provides a powerful framework for analyzing and predicting human information-seeking behavior. Studies have been done to extend the theory and try to understand people's information foraging behaviors in different contexts~\cite{sandstrom1994optimal, lawrance2010programmers, dwairy2011application, vigo2013challenging}. However, few studies have investigated ways to explicitly support people's information foraging processes. Vattam et al. \cite{vattam2011foraging} developed an information-processing model of seeking bio-inspiration in online information environments and proposed an approach for enhancing these environments by augmenting individual information resources with conceptual models, to make finding relevant information more efficient. They divided the inspiration foraging process into two main stages: \textit{within-patch foraging} and \textit{between-patch foraging}, and supported each stage differently to cater to foragers' informational needs. Taking inspiration from these works, Flowy took a novel approach to applying information foraging theory to the UX design inspiration domain. The design considerations in Flowy reflect a human-centered, intelligence augmentation approach of AI adoption for supporting UX designers' information foraging processes of inspirations in early design stages.

%% file: sections/3-formative-study.tex
\section{Formative Studies}

The concept of user flow, while long-standing in design practices~\cite{Kaplan_2023, idf_user_flow, deaton2003elements}, has been understudied in academic research. User flows are often represented using simple flow diagrams and are reflected in various design artifacts, including user journey maps, UI widget flows, and UI prototypes of various modalities~\cite{Kaplan_2023, idf_user_flow}. Despite their widespread use in practice, little empirical research has been done to understand UX designers' practices regarding user flows. Additionally, we observed that most AI tools designed to support UX work primarily focus on generating individual UI screens, neglecting the broader context of user flows in the design process. Consequently, our formative study aimed to explore the following research questions to fill this gap:

\begin{enumerate}
    \item \textbf{RQ1:} What feedback do designers have towards existing prompt-to-UI AI design support tools?
    \item \textbf{RQ2:} How do UX designers manage user flows using different abstractions and design artifacts such as user journey diagrams, UI widget flows, and high-fidelity UI screens?
\end{enumerate}

Our goal is to understand user flow as a unique aspect of UX design beyond static UI design, and further identify opportunities for naturally integrating AI support into UX design workflows. We recruited five professional UX designers through online advertising and word of mouth. A description of the participant demographics is shown in Appendix \ref{tab:formative-demo}. We conducted a semi-structured, one-hour interview with each participant. In our interviews, we used \textbf{retrospective analysis}, a common method to reconstruct the behaviors, rationales, and emotions of the study participants for the recorded events~\cite{russell2014looking}. We asked participants to refer to digital design and research files from a previous design project while explaining the behavior, rationale, and ideas in the design processes. The interview questions were organized into three sections:

\begin{enumerate}
    \item Understanding the participant's previous UX design project;
    \item Practices to manage user journeys, user flows, UI widgets, and high-fidelity prototypes in the design project;
    \item Testing out a representative existing AI-powered prompt-to-UI generation tool, Uizard\footnote{We selected Uizard (https://uizard.io/) as one of the earliest tool in this space. It shares the same core features and very similar experiences with alternatives such as V0 by Vercel (https://v0.dev/) and Galileo (https://www.usegalileo.ai/, not yet available during the time of our study).}.
\end{enumerate}

Each participant was compensated \$15 for their time\footnote{The study protocol was reviewed and approved by the Institutional Review Board
(IRB) at our institution.}. The interview notes were analyzed using affinity diagramming~\cite{pernice_2018} by four of the authors to identify common themes in the responses of the participants. The subsequent sections detail the insights derived from our affinity diagramming analysis.

\subsection{Analysis Findings}
\subsubsection{Finding 1: Existing prompt-to-UI tools provide limited utility for UX design}
\label{sec:formative-finding-1}
    During the interviews, all designers confirmed our hypothesis that existing prompt-to-UI tools are too focused on individual UI screens. Thereby, these tools miss the broader context of user flows vital for UX design, as unanimously noted by the participants (P1–5)~\cite{norman1998definition}. Most participants did not continue to improve on the prompts and generate UI screens after the first round of interaction with Uizard, finding the generated results of little help. P4 reported with frustration that they are \textit{``not sure about the tool’s (Uizard's) jobs to be done... it is very far from current UX designers’ practices''}. P3 emphasized that in UX design \textit{``you are not designing for \textbf{aesthetics}, but for \textbf{functionality}''}, explaining that Uizard cannot support realistic UX design tasks beyond the visual aspects of individual UI screens. With the proliferation of design systems~\cite{churchill2019scaling} that contains pre-defined UI components for UX designers to use, designing high-quality visual UI screens is no longer a major challenge for UX designers~\cite{lu_bridging_2022}, further limiting existing AI tools' value for UX design.
    
    
    Designers identified three main gaps between prompt-to-UI AI tools and realistic UX workflows. First, prompt-to-UI AI tools generate overly generic and simplistic results that cannot meet the requirements of real-world projects. P2 felt that the AI model ``does not understand the complexity that a screen holds'', while P3 mentioned that the AI generated results \textit{``can be helpful only in generic industries''}. The second gap is that no rationale of design decisions are provided with the generated UI results, leaving designers without the insights needed to leverage these outputs as meaningful inspiration. P1 highlighted that in UX inspirations, ``it's really important to understand the rationale behind the design scenarios and the selection of design patterns''. Such rationales contain practical value in communications when designers need to justify their design decisions to teammates and stakeholders (P1, P3, P4). In contrast, examples from existing products, like the ones featured on Mobbin, have been tested with real users, and thus serve as better, more trustworthy inspirations (P1, P4). Lastly, integration with organizational design systems presents a challenge. Due to the widespread expectation that professional UX designers need to work with their organizations' design systems, the inability of current AI tools to accommodate these systems effectively forces designers to engage in almost as much effort as would be required to develop a new prototype from scratch (P1, P3, P4).
    

\subsubsection{Finding 2: User flow management is highly adaptive and project-specific}
\label{sec:formative-finding-2}


Regarding RQ2, designers reported that they adapt or invent many forms of representations and artifacts based on the needs and requirements of individual projects to visualize, understand, and manage user flows (P1, P3, P4, P5). Basic flow diagrams are versatile and functional enough in most cases (P2, P3, P4). UX designers often also need to manage other aspects of the system associated with user flows, such as the features that support each step of the user flow (P4) and the accompanying data flow (P5). P4 mentioned a specific intermediary design artifact named ``breadboard''~\cite{singer_breadboard} they found very useful to manage features in user flows, while recognizing that it is not yet well-known in the UX community. P5's team specifically ``invented'' an artifact called ``receipe'' to list and manage the changing \textit{data flow} across multiple UI screens in a previous project. Meanwhile, P3 pointed out that deliberate management of user flows is only necessary when they are complex and not self-explanatory. In many cases, \textit{``user flows are often directly encapsulated in high-fidelity prototypes''}, with \textit{``no specific abstractions or artifacts for them''} (P3). In general, designers' adaptive management of user flows echos previous research findings that UX methodologies are ``more like mindsets'' than formal procedures to follow~\cite{gray_its_2016}. It helps designers to make sense of the problem domain, which is critical in practical UX design, but unsupported by existing prompt-to-UI tools.

\subsubsection{Finding 3: UX designers use design patterns to understand the design space and scaffold decision-making}
\label{sec:formative-finding-3}

An unexpected finding is the value of \textit{design patterns} in the creation of UX prototypes (P1, P3, P4, P5). The term ``design pattern'' is loosely-defined in UX. Based on our interviews and analysis of previous work~\cite{Tankala_Joyce, nguyen2018deep, tidwell2010designing, neil2014mobile}, we refer to design pattern as \textit{``common ways'' of designing static layout and dynamic interaction across screens}. For professional UX designers, patterns are exemplified through individual UX examples. Design patterns help designers comprehensively understand the potential design space. They scaffold the UX designers' decision-making process by providing rationales for different design choices and connecting design decisions to effects on user experiences. P3 mentioned that ``it's really important to understand the rationale behind the design scenarios and the design patterns to adopt others' designs''.  

We observed that UX designers' search of user flow inspirations greatly resembles the \textit{information foraging} process~\cite{pirolli1999information}. Designers look through the user flow prototypes of existing products, (\textit{find}), implicitly reason about the design scenario and the rationales behind selecting specific design patterns (\textit{evaluate}), and collect those that best fit the design problem at hand (\textit{aggregate}). During our interviews, all participants showed an instinct to reason about the rationales behind the design decisions in example user flow prototypes. Carrying these rationales forward, designers also use them to justify their design decisions to collaborators and stakeholders.


A major challenge is that design pattern knowledge is acquired primarily through experience and deliberate learning over time. For example, when creating prototypes in tools like Figma, designers are often constrained in their ability to explore a broad array of user flow examples. This makes it difficult for them to comprehensively understand the design space and the potential solutions offered by various design patterns. This constraint can lead to locally optimal design decisions based on the narrow range of examples they happen to choose, potentially resulting in design fixation~\cite{}. To account for this, UX designers need to consult design pattern analysis articles on websites like Pencil \& Papers\footnote{https://pencilandpaper.io/articles/user-experience/user-experience-pattern-analysis/} and Built for Mars\footnote{https://builtformars.com/} to learn to design specific UX features (P1, P4). P1 also mentioned that \textit{``as a grew more senior, I accumulate more examples in their head, so I see alternative pattern whenever looking a specific example''}.


    

    



%% file: sections/4-design-goals.tex
\section{Design Goals}
\label{sec:design-goal}
Our formative study revealed a key insight about current UX design processes: with the widespread use of pre-defined components in companies' design systems~\cite{frost_atomic_2016, churchill2019scaling}, the bulk of UX design work has shifted to \textit{adopting the most suitable design patterns} to the design scenario at hand. 

It became evident to us that designers can benefit more directly from intelligent support in understanding the design space, rather than being provided with AI-generated UI screens \hyperref[sec:formative-finding-3]{\textit{(Finding 3)}}. Thus, we designed our prototype, Flowy, with the following design goals:

\textbf{DG1: Support scaffolding design decision making and accelerating information foraging of UX inspirations.}
\label{sec:design-goal-1}
Instead of adopting an approach to generate user flows and using a similar paradigm to prompt-to-UI generation, we designed Flowy to explicitly support designers' existing cognitive processes in inspiration foraging. We observed that UX design processes and decisions are often very complex and context-specific \hyperref[sec:formative-finding-2]{\textit{(Finding 2)}}. It is almost impractical for designers to prompt an AI model with \textit{sufficient} design scenario context to generate useful UX flows that matches their specific project context \hyperref[sec:formative-finding-1]{\textit{(Finding 1)}}. Based on our formative study \hyperref[sec:formative-finding-3]{\textit{(Finding 3)}}, we designed Flowy to support designers' current practices of learning design patterns through user flow examples. Flowy aims to scaffold UX designers' foraging process by helping them \textit{identify} design patterns they can potentially adopt and the \textit{design considerations} to make when deciding whether to adopt a pattern in their specific scenario. This approach allows designers to make \textit{more informed} design decisions supported by evidence, while also accelerating the decision-making process.

\textbf{DG2: Prioritize design patterns regarding user flows and user interactions over individual, static UI screens.}
\label{sec:design-goal-2}
Static UI screens alone provide limited information about the overall user experience and the flow of interactions within a product. In reality, designers work with dynamic, interconnected user flows that define how users navigate through a product and complete tasks \hyperref[sec:formative-finding-1]{\textit{(Finding 1)}}. By providing information on design patterns for user interactions and user flows, we can more closely align our support with the challenges designers face in their everyday work. Providing guidance and inspiration at the user flow level helps designers understand how different screens and interactions fit together to create a cohesive user experience. 

\textbf{DG3: Provide high-quality design pattern insights connected to individual examples.}
\label{sec:design-goal-3}
In designing AI-powered tools to support UX designers, providing high-quality design pattern information and avoid hallucination is crucial. Only by providing high-quality, professional-grade design pattern information, can we ensure that designers receive reliable guidance and inspiration \hyperref[sec:formative-finding-3]{\textit{(Finding 3)}}. Furthermore, it is important to recognize that design patterns and specific user flow examples serve different but complementary purposes. Design patterns represent aggregated knowledge and best practices, offering a high-level understanding of common solutions~\cite{Tankala_Joyce, nguyen2018deep, tidwell2010designing, neil2014mobile}. On the other hand, individual user flow screenshots showcase specific design implementation details, such as component usage, placement, and interaction details. We design Flowy to provide both pattern-level information and concrete examples, so that designers can gain a comprehensive understanding of the design space and make informed decisions tailored to their specific context.

\textbf{DG4: Provide easy-to-consume design pattern information to accelerate design decision making.}
\label{sec:design-goal-4}
It is essential to provide UX designers with design pattern information that is easy to understand and apply. Overwhelming designers with complex or overly detailed information can hinder their ability to quickly grasp key insights and make informed decisions~\cite{koch_may_2019, kim_effect_2023}. This is particularly important in design ideation, where designers often need to explore and evaluate a wide range of examples. By presenting design patterns in a clear, concise, and easily digestible format, we can help designers rapidly process and integrate this knowledge into their work. This can also enable designers to more flexibly manage user flows and adapt their design decisions based on the specific project context \hyperref[sec:formative-finding-2]{\textit{(Finding 2)}}.





%% file: sections/5-system-design.tex
\section{Flowy Design \& Implementation}
In our Flowy prototype, we implemented a design pattern annotation pipeline that utilizes the latest multimodal AI with the high-quality user flow prototype dataset Mobbin. We take design pattern analysis for each user flow prototype, then propagated the annotation results to our web app's frontend.

\begin{figure*}
    \centering
    \includegraphics[width=\linewidth]{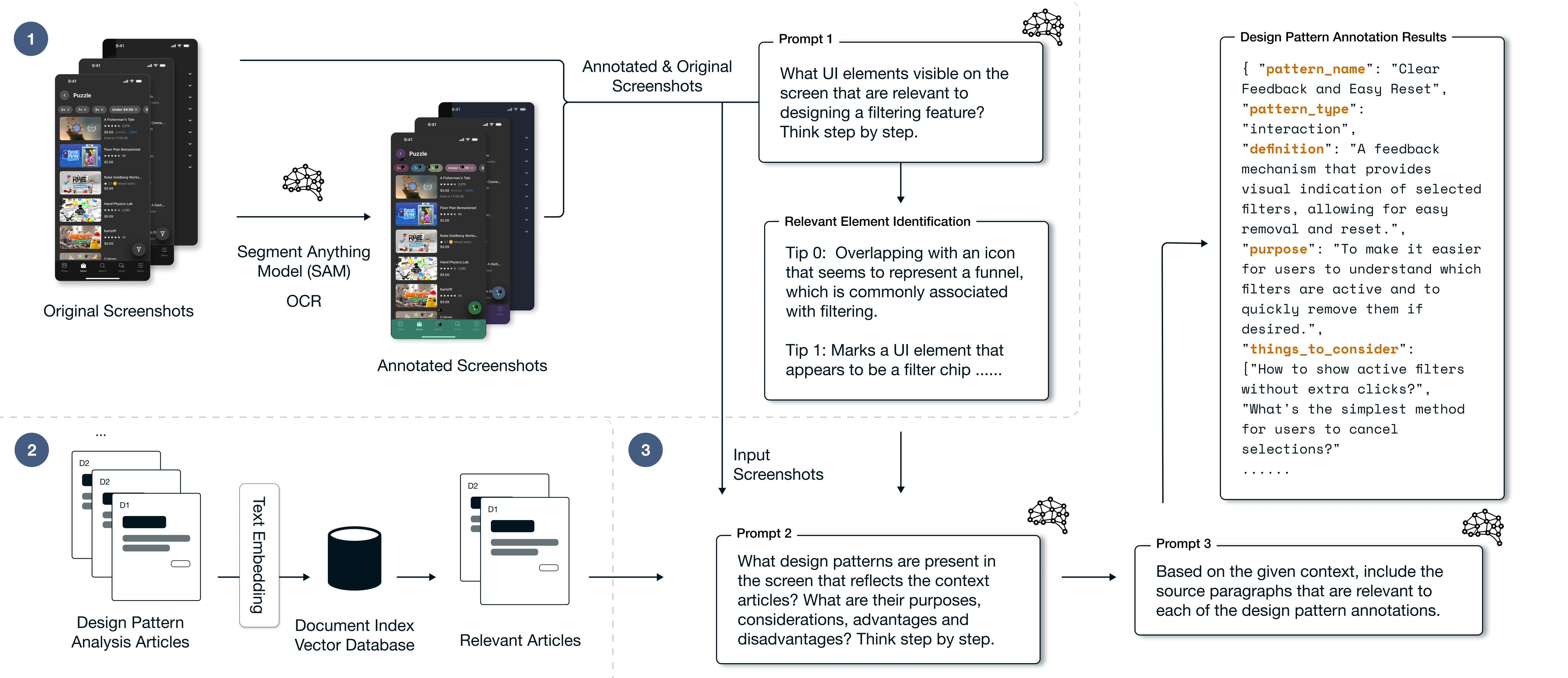}
    \caption{Flowy's backend design pattern annotation pipeline. The pipeline utilizes Set-of-Mark prompting (1) and Retrival-Augmented-Generation (2). It takes original screenshots of user flow examples (1), which are passed through a Segment Anything Model (SAM) and Optical Character Recognition (OCR) to identify relevant UI elements. Relevant design pattern analysis articles are retrieved from our curated knowledge base, stored in a vector database (2). The screenshots and retrieved articles are input to a multimodal AI model to generate design pattern annotations, including the pattern name, purpose, design considerations, etc. (3). Prompts are simplified to improve readability yet still highlight the core utilities.}
    \label{fig:annotation_pipeline}
    \Description{The figure depicts an annotation pipeline for generating design pattern information from app screenshots. The pipeline consists of three main stages: (1) Inputting original screenshots from user flow examples, which are processed to identify relevant UI elements, (2) Retrieving relevant design guideline articles from a curated knowledge database, and (3) Inputting the screenshots and articles into a language model to generate detailed design pattern annotations. The generated annotations include key information such as pattern name, purpose, advantages and disadvantages, and design considerations.}
\end{figure*}

\subsection{Design Pattern Annotation Pipeline}
Our annotation pipeline aims to provide high-quality, professional-grade design pattern information that is easy to consume \textit{(\hyperref[sec:design-goal-3]{DG3}, \hyperref[sec:design-goal-4]{DG4})}. Thus, we designed a Set-of-Mark (SoM) prompting module to anchor annotations onto specific visual UI elements on screens, together with a Retrieval-Augmented Generation (RAG) module to ground our annotation in existing high-quality UX design pattern analysis articles. Generally, we used state-of-the-art large multimodal models to annotate design pattern information on user flow prototypes.\looseness=-1

\subsubsection{Visually-Anchored Annotation Through Set-of-Mark Prompting}
\label{sec:som-prompt}
We employed Set-of-Mark (SoM) prompting, a method for improving the visual grounding abilities of large multimodal models~\cite{Yang_Zhang_Li_Zou_Li_Gao_2023}, to associate design pattern annotations with visual UI elements on the screenshots (Section 1 in Figure \ref{fig:annotation_pipeline}). SoM is typically done by marking images with main object region masks from an object detection model and associating each mask with a numeric ``tip''. The original and marked images are then fed into a multimodal AI model. In our project, using SoM enables us to relate design pattern annotations with associated UI elements in user flow examples, improving the ease of consumption for designers \textit{(\hyperref[sec:design-goal-3]{DG3}, \hyperref[sec:design-goal-4]{DG4})}.

\subsubsection{High-Quality Design Pattern Information With Retrieval Augmented Generation}
\label{sec:rag-prompt}
To ensure the quality of our annotation results, the research team manually curated 54 high-quality articles on UX design pattern analysis through collective discussion and suggestions from professional UX designers, as our design pattern knowledge base \textit{(\hyperref[sec:design-goal-2]{DG2}, \hyperref[sec:design-goal-3]{DG3})}. We used Retrieval-Augmented Generation (RAG)~\cite{lewis2020retrieval}, a common technique in knowledge-intensive tasks to supplement large multimodal models with domain-specific information. We employed RAG to retrieve relevant articles from our curated knowledge base based on the similarity between the article content and the input user flow example (Section 2 in Figure \ref{fig:annotation_pipeline}). The retrieved articles were then used as additional context in our prompt to guide the model in generating accurate and informative design pattern annotations.

\subsubsection{Specific Features for Efficient UX Inspiration Foraging}
\label{sec:specific-foraging-support-features}
We took inspiration from past research on inspiration foraging supporting tools~\cite{vattam2011foraging, goel2012cognitive} and specifically designed our annotation task around the \textit{within-patch foraging} and \textit{between-patch foraging} state of designers \textit{(\hyperref[sec:design-goal-1]{DG1})}. In within-patch foraging, the forager gains net information by consuming the content of the information patch. To support designers' net information gain of design patterns in user flow examples \textit{(\hyperref[sec:design-goal-4]{DG4})}, on top of design pattern definitions and descriptions, we also defined key aspects of rationale behind the adoption of design patterns based on previous work on design patterns~\cite{tidwell2010designing, neil2014mobile}, namely purposes, advantages, disadvantages, and design considerations and generated such information based on the RAG-retrieved design guideline articles. Such information adds values to designers by providing them more granular details in the context, trade-offs, and considerations behind the use of specific design patterns. During \textit{between-patch foraging}, the forager tries to find the next best information patches to consume. Thus, we provide screenshots of related designs adopting similar design patterns to help designers identify the other patches of interest (see visualized illustration in Section~\ref{sec:frontend}).

\subsubsection{Prompt Structure}

The general prompt structure of our design pattern annotation pipeline consists of three main prompts (Figure \ref{fig:annotation_pipeline}). In the first prompt, we ask the large multimodal model to visually identify and explain UI elements that are relevant to the design pattern of interest. This prompt was designed to \textit{ground} the subsequent annotations in accurate understandings of each UI element's function and reduce the model's hallucination. Then, the explanation of relevant UI elements were combined with retrieved professional UX pattern analysis articles (Section \ref{sec:rag-prompt}), to prompt the model to annotate the purposes, considerations, and advantages/disadvantages of a design pattern example. In the last prompt, we asked the model to reflectively select the most relevant sections of UX pattern analysis articles for each annotation, to provide users with accurate source information as references.

\subsubsection{Annotation Pipeline Implementation Details}

To implement SoM prompting, we first performed UI element detection using the Segment Anything Model (SAM)~\cite{kirillov2023segment}, which generates fine-grained segmentation masks over UI elements on screenshots. To mitigate over-segmentation on text content, where texts on screenshots are divided into overly small pieces, we used Optical Character Recognition (OCR) to identify text regions and filtered out SAM segmentations where the Intersection over Union (IoU) with any text region exceeded 0.55, a threshold empirically set through experimentation. Additionally, tiny segmentations constituting less than 0.5\% of the screen area were filtered out. The resulting segmentation masks, their numerical identifiers, and the original screenshots were then input into the large multimodal model (Figure \ref{fig:annotation_pipeline}), following the approach described in the original SoM paper.

To implement our RAG component, we used cosine similarity between the article content embeddings and the current product's feature name to retrieve design pattern analysis for the specific feature\footnote{In our curated knowledge base, design pattern analysis articles are mostly written around specific product features.}. Specifically, we used LangChain and ChromaDB to implement this component. Then, we added the retrieved article's original text, together with the original and marked screenshots from our SoM prompting module, into the multimodal AI model to annotate for the design pattern information on UI screenshots.

We selected GPT-4V as the multimodal AI model for Prompt 1 and Prompt 2 (Figure \ref{fig:annotation_pipeline}). Prompt 2 and 3 were initially a same prompt. However, in our experimentation, we empirically observed that GPT-4V can seldom correctly retrieve the relevant paragraphs from source pattern analysis documents, yet Claude-3-Opus was able to reliably perform this task. As a result, we separated Prompt 3 from 2 and specifically used Claude-3-Opus for Prompt 3. Our prompt structure adopts the Chain-of-Thought~\cite{wei2022chain} and prompt chaining~\cite{wu2022ai} techniques to enhance the generation pipeline's performance. We annotate all user flow prototypes in advance and statistically store them in our web app, to ensure information consistency and avoid long inference time.

\begin{figure*}
    \centering
    \includegraphics[width=\linewidth]{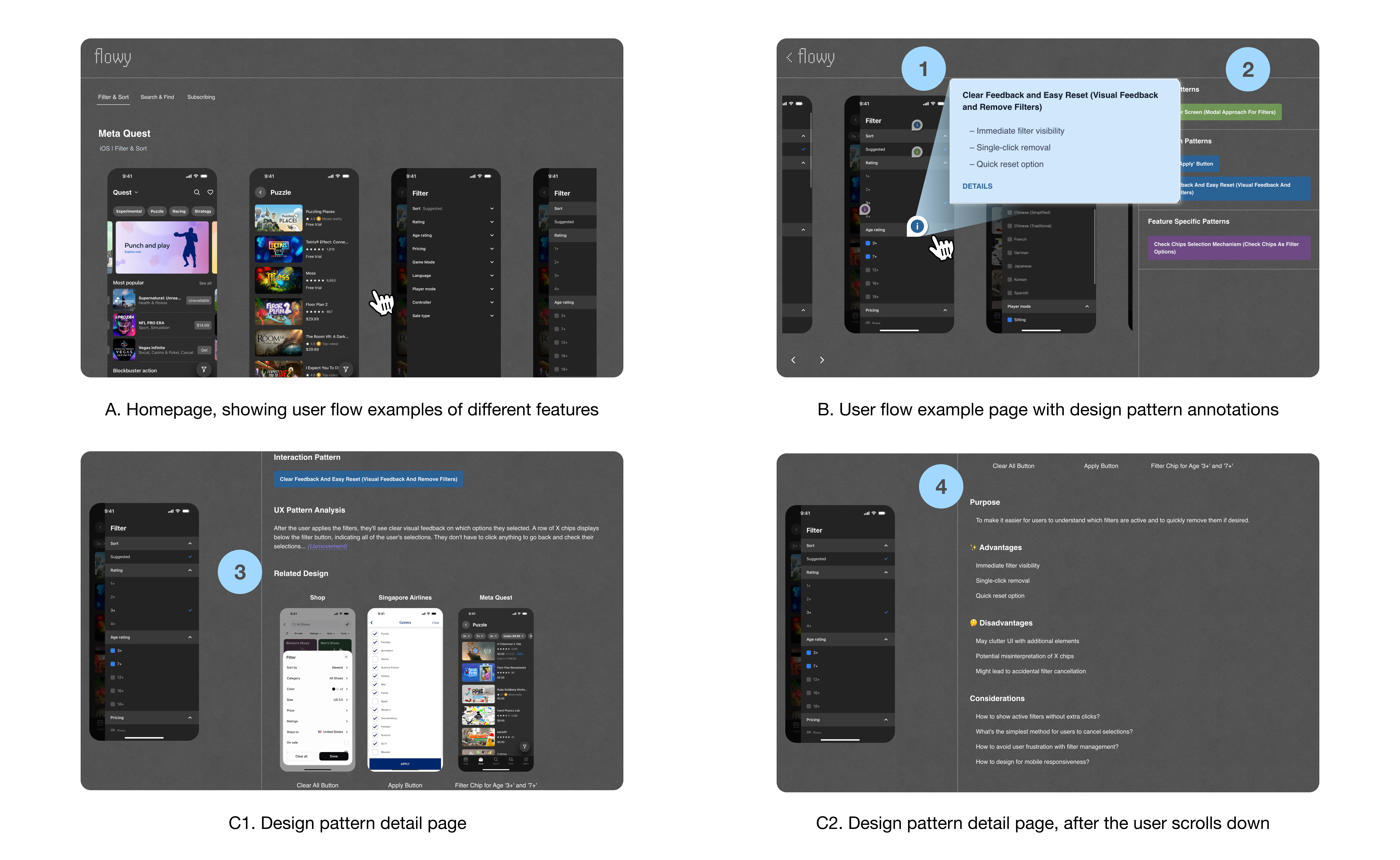}
    \caption{Flowy's frontend design offers a multi-layered, flexible way to explore design examples and associated design patterns at different granularity. The homepage (A) showcases user flows from existing products. Selecting a flow opens an interactive annotated view (B), with key elements highlighted and briefly explained on hover. Clicking an annotation reveals a dedicated page (C1) with detailed pattern analysis and related design screenshot examples for side-by-side comparison (C2).}
    \label{fig:frontend}
    \Description{The image shows 4 screenshots demonstrating the user interface and flow of a tool called Flowy. Image A shows the homepage with examples of user flows from different apps. Image B shows a user flow page with design pattern annotations. Image C1 shows a detailed design pattern page, and image C2 shows that same page scrolled down to reveal additional information and example screenshots related to the selected design pattern. The overall figure illustrates how Flowy enables exploring UI/UX design examples at varying levels of detail, from high-level flows down to specific design patterns.}
\end{figure*}

\subsection{Frontend Web Design \& Implementation}
\label{sec:frontend}

To support the efficient consumption of design pattern information through specific design examples \textit{(\hyperref[sec:design-goal-1]{DG4})}, Flowy's frontend adopts the Visual Information-Seeking Mantra: overview first, zoom and filter, then details-on-demand~\cite{north2000snap}. As illustrated in Figure \ref{fig:frontend}, after a user selects a user flow example on Flowy's homepage (Screen A), they enter the page displaying annotations for this user flow (Screen B). The icon tags (1) on each UI screenshot, indicating user pattern annotations and their associated UI elements, together with the list of design pattern annotations (2) on the right, provide an \textit{overview} of the available annotations. The user can hover over annotations they are interested in and see a popup window with concise, bite-sized summaries regarding this design pattern in a bullet list \textit{(zoom and filter)}. If the user is interested in learning further, they can click on the annotation popup to go to the annotation detail page (Screen C1, C2), where related designs (3) and more details regarding this design pattern (4) are provided \textit{(details-on-demand)}. This multi-level, interactive design allows users to quickly grasp the overall design patterns present in a user flow example while enabling them to selectively dive deeper into patterns of interest, facilitating efficient design inspiration foraging.

As discussed in Section \ref{sec:specific-foraging-support-features}, we support UX designers' inspiration foraging process by specifically catering to the \textit{within-patch foraging} and \textit{between-patch foraging} stages. The detailed information (4) regarding a design pattern increases a designer's net information gain during \textit{within-patch foraging}, while the ``Related Design'' section (3) helps a designers to identify potential patches to look through next.

The frontend of Flowy was implemented using Next.js and Tailwind CSS. As a prototype, in Flowy, we included 31 unique, high-quality user flow examples for three product features, namely Filter \& Sort, Search \& Find, and Subscribing. On Screen C1, the ``Related Design'' section (3) was implemented by ranking the cosine similarities between pattern definitions in user flows for the same product feature.


%% file: sections/6-user-study.tex
\section{User Study}

We conducted a user study with eight UX designers of varying experience to evaluate the Flowy prototype\footnote{The study protocol was reviewed and approved by the Institutional Review Board (IRB) at our institution.}. Among the eight participants, one participant is a student in UX while the rest are all professional UX designers. A table describing the participant demographics
is shown in Appendix \ref{tab:user-study-demo}. Each study session lasted one hour, and participants received \$15 USD as compensation for their time. Our study aimed to answer the following research questions:

\begin{itemize}
    \item \textbf{RQ3}\footnote{We continued the RQ numbering to differentiate from research questions in our formative study.}\textbf{:} How might Flowy support UX designers' ideation process in creating UX prototypes?
    \item \textbf{RQ4:} How might Flowy fit into existing UX design workflows and provide support?
    \item \textbf{RQ5:} What challenges do UX designers encounter when using Flowy for UX ideation support?

\end{itemize}

\subsection{Study Design}
Each study session was conducted online and lasted approximately one hour. The sessions consisted of two main activities: system feedback (20-25 minutes) and a mock design task (20-25 minutes). First, after obtaining informed consent from the participants, the researchers introduced them to Flowy and provided a high-level explanation of its key features. Then, each participant independently navigated through the interfaces of Flowy as a potential user of the system. The participants were asked to provide feedback regarding the design pattern annotations, focusing on how the annotation information might help or hinder their everyday design tasks. Collecting feedback on the content of design pattern annotations was crucial to ensure that the information provided by Flowy was relevant, accurate, and helpful for UX designers in their real-world projects.

Afterwards, each participant was tasked with designing a filter feature inside an existing Figma prototype for a mobile electronics shopping app. This mock design activity aimed to simulate a realistic UX design scenario and assess the effectiveness of Flowy in supporting practical design challenges within participants' actual workflows. The researchers encouraged the participants to refer to Flowy during the design task while also inviting them to use other tools they would typically employ in their natural workflows. During the design task, participants were instructed to think aloud, enabling researchers to gain a deeper understanding of their thought processes~\cite{mcdonald2012exploring}. Researchers asked follow-up questions throughout the study for clarification. At the end of the session, participants completed a post-study questionnaire, which gathered feedback regarding the usability and usefulness of the Flowy prototype, as well as its potential integration into their existing UX design workflows.


\begin{figure*}
    \centering
    \includegraphics[width=0.8\linewidth]{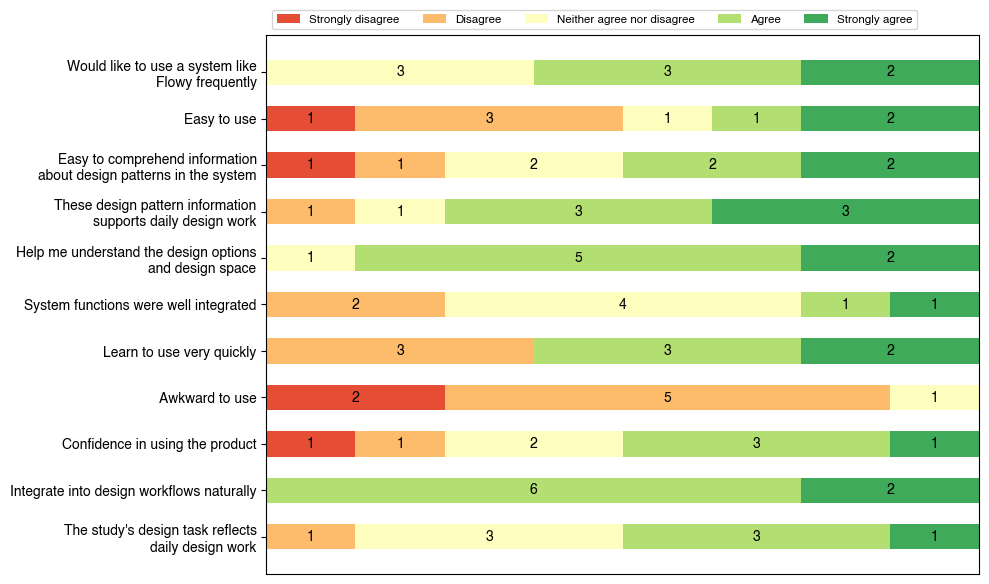}
    \caption{Questionnaire results from our user study.}
    \label{fig:questionnaire}
    \Description{Figure 4 shows questionnaire results from the user study, with various statements about the Flowy system on the y-axis and the number of participants who strongly disagree, disagree, are neutral, agree, or strongly agree with each statement represented by colored bars. The results indicate that most participants would like to use a system like Flowy frequently, found it easy to use, and thought the design pattern information provided was easy to comprehend and helped them understand design options and the design space. The majority also felt the system functions were well integrated, they learned to use it very quickly, and agreed it integrates into their design workflows naturally. Overall, the questionnaire responses suggest participants had a positive experience with Flowy and found it helpful for understanding relevant design patterns and options to support their UX design work. }
\end{figure*}

\subsection{Key Insights}

Four of the researchers collaboratively analyzed the notes take during the user studies as a group qualitatively using affinity diagramming~\cite{pernice_2018}. The goal is to identify common themes in the responses of the user study participants. The questionnaire responses collected at the end of user study sessions are also visualized in Figure \ref{fig:questionnaire}. In this section, we present the key insights from our annotations with reference to the quantitative measures in the questionnaires.

\subsubsection{Insight 1: Flowy can accelerate inspiration foraging through more informed, evidence-based UX design decisions.}

Our participants generally found value in Flowy's features for scaffolding their considerations in UX design decision making and accelerating their inspiration foraging in the mock design activity. This is reflected in questionnaire responses (Figure \ref{fig:questionnaire}) regarding Flowy's high value on understanding the design space and good potential to naturally integrate into designers' workflows. Confirming our design goal \textit{\hyperref[sec:design-goal-2]{(DG2)}}, our participants highlighted that design decisions at the user flow level as a critical yet often overlooked aspect of UX design (P1, P4, P7). P1 highly appreciated the ``Related Design'' section's examples (Figure \ref{fig:frontend}, element 3), which helped her to directly compare similar design patterns and identify more promising inspirational examples. P7 specifically said that for the design pattern analysis, \textit{``if you do not do it for me, I will have to do it myself''}.

Participants found Flowy's annotations on the flow example page (Figure \ref{fig:frontend}, element 1) helpful in guiding their attention onto important UI elements. P4 and P5 noted that Flowy's annotations help prevent overlooking relevant UI elements in user flow examples, a common issue when looking through many user flow examples. In addition, P2, P4, P5, and P7 mentioned that Flowy's design pattern annotations can serve as positive examples of \textit{design rationales}, which can greatly help UX designers in justifying their design decisions in intra- and inter-team communications. This echos findings in previous empirical research of UX practices~\cite{lu_bridging_2022}. P8 said that \textit{``(Flowy) can be particularly helpful for UX beginners when they work on new features they are not familiar with''}, a sentiment also shared by P1 and P5. In general, our participants would want to use Flowy's features in their everyday work and mostly think it will easily integrate into their natural workflows, as reflected in the post-study questionnaire responses (Figure \ref{fig:questionnaire}).

\subsubsection{Insight 2: Designers desire design pattern annotations beyond user flows, including product, business, and development considerations.}

Our UX design pattern annotations primarily focused on user flow and interaction patterns as a result of our knowledge base articles' focus. However, our participants also highlighted that in practice, their design decisions often also involve considerations for their products' goals, business objectives, and technical development limitations. These further considerations often cannot be directly inferred from the visible UI screens and UX prototypes. However, they can significantly help designers in evaluating design patterns' adaptability to their own design scenarios. P4 analyzed that \textit{``intuitively I have a gut instinct of why the designer did this, and it's not really mind-blowing to read the advantages and disadvantages (of a design pattern)... what would be really interesting that I am curious on, is more product-specific context, ... why this design decision relates back to the app store design specifically''}. Similar opinions are also shared by P2 and P6 during the sessions.

This uncovered the fact that UX design decisions often are numerous determined by factors beyond user flows and experiences. We summarized the factors mentioned by designers into three main categories: target user groups, (P1, P2, P5, P7), product decisions (P2, P5, P7), and development constraints (P2, P5, P6). 
UX designers often have to base their design decisions of user flows and multi-screen component flows on careful trade-offs between these main factors. As a result, the UX ideation process can be very cognitively demanding. We observed that during our study, P2, P4, P5, and P7 all implicitly quickly ignored overly complex user flow examples and information that is not straightforward on screen to conserve cognitive resources. This inspired us to create more concise, straightforward, and specific design pattern annotations in future iterations of Flowy.


\subsubsection{Insight 3: Designers prioritize visual comparative learning of design patterns.}

Interestingly, during the study sessions, the majority of participants spent the most time trying to understand design patterns by using the ``Related Design'' section in Flowy, where design examples adopting similar design patterns are displayed side-by-side (P1-4, P6, P8). Six out of eight participants (P2-4, P6-8) explicitly mentioned that they preferred comparing different user flow screenshots to reading textual design pattern analysis, although they acknowledge the latter can provide design rationale references. More specifically, P4 mentioned that they \textit{``do not want to read the text''}, despite our efforts to make textual information short and concise. We assume that this, combined with the fact that Flowy's pattern analysis is mostly represented as text (Figure \ref{fig:frontend}, element 4), is the main reason behind some participants perceiving Flowy as not easy to use in the questionnaire response. Yet, when P4 was asked what information regarding the design patterns they would find helpful, they describe information with the content and representation very similar to what Flowy has. P4 articulated: \textit{``I don't know, I feel like something just makes me not want to read this... but to be honest, that's also where I am confused in myself that, I understand you guys have these (information).''}\looseness=-1

We believe that this prioritization reflects the \textit{Structural Alignment Theory}, which proposes that when comparing two items, people align their common relational structures to identify similarities and differences~\cite{gentner1997structure}. This process has been applied to the abstraction of common principles and design ideation~\cite{dahl2002influence, chan2018best, gassmann2008opening, ozkan2013cognitive}. For our participants, learning design patterns through visual comparisons of similar design examples was more effective than reading textual explanations, even though the latter were also intentionally designed to be concise and easy-to-consume. P6 explained that they only used textual information as a fallback when they cannot intuitively understand visceral information. This key insight provides us with a future design implication to prioritize visual comparative annotation of design patterns, over textual explanations.


\subsection{Design Implications}

We generalized our study findings to derive key design implications for AI-powered UX ideation tools like Flowy. These implications highlight the needs and preferences of UX designers when using such tools in their design workflows.

\begin{enumerate}
    \item Provide concise, straightforward, and precise design pattern annotations to reduce designers' cognitive load;
    \item Expand design pattern analysis beyond user flows and interactions, to include relevant product goals, business objectives, and technical development considerations.
    \item Prioritize visual, comparative learning of design patterns between screenshot examples over providing textual explanations.
\end{enumerate}

Based on these design implications, we plan to iterate on the Flowy prototype to better align with the needs and expectations of UX designers. We believe that these implications can also serve as valuable guidelines for future research in creativity support for UX design.





    

%% file: sections/7-discussion.tex
\section{Discussion}

\subsection{Designerly Understanding of UX Protoypes with Multimodal AI}

While some previous research in AI has tried to integrate professional UX designers' considerations into computational deep learning models~\cite{schoop_predicting_2022, li_screen2vec_2021, huang_swire_2019}, but most existing work in AI takes end users as the target group for UI/UX related tasks~\cite{wang_enabling_2023, wang_screen2words_2021, yan2023gpt}. Much of supposedly designer-centric AI research seems to be disconnected with designers' latest practices, focusing on concepts such as aesthetics that are not grounded in design practitioners' main considerations. During our design of Flowy, we empirically observed that current large multimodal AI models can only recognize basic UI elements and their purposes (end-user-centric), but are limited in understanding the design rationale and considerations behind design decisions (designer-centric). 

We assume that these phenomena reveal the current gap and many potential directions in AI support for UX design. We call for the empirical understanding of two important aspects of UX design for future AI research: UX is much more than UI, and UI screens often embed more design considerations than a good composition of visual UI components~\cite{lu2024ai}. Coming up with good composition of UI layouts is considered as UI design and only part of the goals for UX designers~\cite{norman1998definition}. The more important end goal is to construct user flows and interactions based on established best practices and design patterns, fitting the UI layouts, interactions, and flows to users' goals, business values, and development resources.

\subsection{Transforming UX Design Practices: Supporting Evidence-Based Decisions, Automating Implementations}

In our formative studies, we found that the widespread use of design systems and pre-defined design component libraries shifted the essence of UX design tasks to adopting the most suitable design patterns to the design scenario at hand. Many existing generative AI tools for UX aim to automate the creation of UI screens through a prompt-to-UI paradigm. Our effort to prototyping Flowy represents a complementary approach to support UX designers in making evidence-based design decisions through design pattern annotations on high-quality user flow prototype examples. We argue that by combining these two approaches and carefully design the augmentation and automation roles of AI in different stages of UI/UX design, great future directions can be pursued to transform the current UX design practices. Additionally, by considering latest advancements in streamlining the designer-developer hand-off process~\cite{feng_guis2code_2021, feng_handoffs_2022, feng_understanding_2023}, further potential lies in the general pipeline and paradigm of software design and implementation.

\subsection{Understanding Task Complexity to Design AI Support for Professional Users}

In our formative study, we confirmed our assumption of UX designers' heavy focus on user flows across multiple screens than individual UI screens. We also uncovered the key insight that current UX design processes is more about adopting the most suitable design patterns to the current design scenario, than coming up with visually appealing UI screens. From our user study of Flowy, we discovered that UX designers also work with high complexity in target user groups, business goals, product decisions, and development resources. These findings revealed that UX design processes are much more complex than visible UI screens.

While grounded in the unique characteristics of UX, the principles underlying Flowy's design are generalizable to other creative tasks. By choosing representations with appropriate levels of abstraction (e.g. user flow examples, design pattern knowledge), Flowy allows users to make sense of the solution space at various conceptual levels. Additionally, by helping users navigate through connections and affinities during the foraging process, Flowy facilitates the exploration and sensemaking of the solution space. These design considerations are particularly relevant to many open-ended tasks characterized by the absence of a single optimal solution, where decision-making requires careful reasoning over a comprehensive understanding of the solution space. Flowy's approach embodies a human-centered, intelligence augmentation approach~\cite{hassani2020artificial} of human-AI collaboration, with the end goal of supporting target users' cognitive processes to improve task efficiency and effectiveness.

Human-centered adoption of AI requires deep knowledge and understanding of the complexity of tasks of professionals and goes beyond automating the creation of visible artifacts~\cite{shneiderman_human-centered_2022}. Our design considerations for Flowy suit many tasks with no single optimal solution, where decision making requires careful reasoning over the solution space. More strategies should be devised for future human-centered AI research as design principles, such as choosing mediums with appropriate levels of abstraction and assisting users in discovering connections and affinities in the foraging process. By focusing on similar key design principles, future AI systems can effectively support professionals in navigating the complex decision-making processes involved in their work, while ensuring that users retain control and maintain a sense of ownership over the creative process.\looseness=-1

%% file: sections/8-conclusion-ack.tex
\section{Limitaiton}
Despite our best efforts to mitigate large-scale multimodal AI's hallucination using Retrieval-Augmented Generation and Chain-of-Thought prompting, we cannot fully avoid design patterns that deviate from the user flow examples. However, we expect this limitation to gradually improve as future multimodal AI model capabilities advance. It is also important to acknowledge Flowy's current limitations in scope and the product features it supports as a prototype. Our evaluation study also has several threats to validity: the representativeness of our participants and tasks may not fully capture the diversity of real-world UX design scenarios. Additionally, the novelty bias associated with using a new tool like Flowy could have influenced the participants' behaviors and attitudes during the study. To address these limitations and gain deeper insights into Flowy's long-term impact on UX design practices, we plan to conduct longitudinal deployments in real-world design settings for future research.

\section{Conclusion}
In this work, we presented Flowy, an AI-powered tool that supports UX designers' inspiration foraging process by supplementing user flow examples with distilled, high-quality design pattern knowledge. Our user study demonstrated Flowy's effectiveness in supporting realistic UX design tasks, with participants finding the visually anchored design pattern annotations helpful for efficiently exploring and understanding relevant design patterns. The design considerations in Flowy, such as multi-level representations and interactive exploration, reflect a human-centered, intelligence augmentation approach to leveraging AI in creative workflows. Our design implications summarized from our user studies point out meaningful considerations for future UX design support tools.


    

%% file: main.bbl

\begin{thebibliography}{78}


\ifx \showCODEN    \undefined \def \showCODEN     #1{\unskip}     \fi
\ifx \showDOI      \undefined \def \showDOI       #1{#1}\fi
\ifx \showISBNx    \undefined \def \showISBNx     #1{\unskip}     \fi
\ifx \showISBNxiii \undefined \def \showISBNxiii  #1{\unskip}     \fi
\ifx \showISSN     \undefined \def \showISSN      #1{\unskip}     \fi
\ifx \showLCCN     \undefined \def \showLCCN      #1{\unskip}     \fi
\ifx \shownote     \undefined \def \shownote      #1{#1}          \fi
\ifx \showarticletitle \undefined \def \showarticletitle #1{#1}   \fi
\ifx \showURL      \undefined \def \showURL       {\relax}        \fi
\providecommand\bibfield[2]{#2}
\providecommand\bibinfo[2]{#2}
\providecommand\natexlab[1]{#1}
\providecommand\showeprint[2][]{arXiv:#2}

\bibitem[\protect\citeauthoryear{Brade, Wang, Sousa, Oore, and Grossman}{Brade et~al\mbox{.}}{2023}]%
        {brade2023promptify}
\bibfield{author}{\bibinfo{person}{Stephen Brade}, \bibinfo{person}{Bryan Wang}, \bibinfo{person}{Mauricio Sousa}, \bibinfo{person}{Sageev Oore}, {and} \bibinfo{person}{Tovi Grossman}.} \bibinfo{year}{2023}\natexlab{}.
\newblock \showarticletitle{Promptify: Text-to-image generation through interactive prompt exploration with large language models}. In \bibinfo{booktitle}{\emph{Proceedings of the 36th Annual ACM Symposium on User Interface Software and Technology}}. \bibinfo{pages}{1--14}.
\newblock


\bibitem[\protect\citeauthoryear{Chan, Dow, and Schunn}{Chan et~al\mbox{.}}{2018}]%
        {chan2018best}
\bibfield{author}{\bibinfo{person}{Joel Chan}, \bibinfo{person}{Steven~P Dow}, {and} \bibinfo{person}{Christian~D Schunn}.} \bibinfo{year}{2018}\natexlab{}.
\newblock \showarticletitle{Do the best design ideas (really) come from conceptually distant sources of inspiration?}
\newblock \bibinfo{journal}{\emph{Engineering a Better Future: Interplay between Engineering, Social Sciences, and Innovation}} (\bibinfo{year}{2018}), \bibinfo{pages}{111--139}.
\newblock


\bibitem[\protect\citeauthoryear{Churchill}{Churchill}{2019}]%
        {churchill2019scaling}
\bibfield{author}{\bibinfo{person}{Elizabeth~F Churchill}.} \bibinfo{year}{2019}\natexlab{}.
\newblock \showarticletitle{Scaling UX with design systems}.
\newblock \bibinfo{journal}{\emph{Interactions}} \bibinfo{volume}{26}, \bibinfo{number}{5} (\bibinfo{year}{2019}), \bibinfo{pages}{22--23}.
\newblock


\bibitem[\protect\citeauthoryear{Dahl and Moreau}{Dahl and Moreau}{2002}]%
        {dahl2002influence}
\bibfield{author}{\bibinfo{person}{Darren~W Dahl} {and} \bibinfo{person}{Page Moreau}.} \bibinfo{year}{2002}\natexlab{}.
\newblock \showarticletitle{The influence and value of analogical thinking during new product ideation}.
\newblock \bibinfo{journal}{\emph{Journal of marketing research}} \bibinfo{volume}{39}, \bibinfo{number}{1} (\bibinfo{year}{2002}), \bibinfo{pages}{47--60}.
\newblock


\bibitem[\protect\citeauthoryear{Deaton}{Deaton}{2003}]%
        {deaton2003elements}
\bibfield{author}{\bibinfo{person}{Mary Deaton}.} \bibinfo{year}{2003}\natexlab{}.
\newblock \showarticletitle{The elements of user experience: user-centered design for the Web}.
\newblock \bibinfo{journal}{\emph{interactions}} \bibinfo{volume}{10}, \bibinfo{number}{5} (\bibinfo{year}{2003}), \bibinfo{pages}{49--51}.
\newblock


\bibitem[\protect\citeauthoryear{Deka, Huang, Franzen, Hibschman, Afergan, Li, Nichols, and Kumar}{Deka et~al\mbox{.}}{2017}]%
        {deka_rico_2017}
\bibfield{author}{\bibinfo{person}{Biplab Deka}, \bibinfo{person}{Zifeng Huang}, \bibinfo{person}{Chad Franzen}, \bibinfo{person}{Joshua Hibschman}, \bibinfo{person}{Daniel Afergan}, \bibinfo{person}{Yang Li}, \bibinfo{person}{Jeffrey Nichols}, {and} \bibinfo{person}{Ranjitha Kumar}.} \bibinfo{year}{2017}\natexlab{}.
\newblock \showarticletitle{Rico: {A} {Mobile} {App} {Dataset} for {Building} {Data}-{Driven} {Design} {Applications}}. In \bibinfo{booktitle}{\emph{Proceedings of the 30th {Annual} {ACM} {Symposium} on {User} {Interface} {Software} and {Technology}}}. \bibinfo{publisher}{ACM}, \bibinfo{address}{Québec City QC Canada}, \bibinfo{pages}{845--854}.
\newblock
\showISBNx{978-1-4503-4981-9}
\urldef\tempurl%
\url{https://doi.org/10.1145/3126594.3126651}
\showDOI{\tempurl}


\bibitem[\protect\citeauthoryear{Deka, Huang, and Kumar}{Deka et~al\mbox{.}}{2016}]%
        {deka_erica_2016}
\bibfield{author}{\bibinfo{person}{Biplab Deka}, \bibinfo{person}{Zifeng Huang}, {and} \bibinfo{person}{Ranjitha Kumar}.} \bibinfo{year}{2016}\natexlab{}.
\newblock \showarticletitle{{ERICA}: {Interaction} {Mining} {Mobile} {Apps}}. In \bibinfo{booktitle}{\emph{Proceedings of the 29th {Annual} {Symposium} on {User} {Interface} {Software} and {Technology}}} \emph{(\bibinfo{series}{{UIST} '16})}. \bibinfo{publisher}{Association for Computing Machinery}, \bibinfo{address}{New York, NY, USA}, \bibinfo{pages}{767--776}.
\newblock
\showISBNx{978-1-4503-4189-9}
\urldef\tempurl%
\url{https://doi.org/10.1145/2984511.2984581}
\showDOI{\tempurl}


\bibitem[\protect\citeauthoryear{Dhariwal and Nichol}{Dhariwal and Nichol}{2021}]%
        {dhariwal2021diffusion}
\bibfield{author}{\bibinfo{person}{Prafulla Dhariwal} {and} \bibinfo{person}{Alexander Nichol}.} \bibinfo{year}{2021}\natexlab{}.
\newblock \showarticletitle{Diffusion models beat gans on image synthesis}.
\newblock \bibinfo{journal}{\emph{Advances in neural information processing systems}}  \bibinfo{volume}{34} (\bibinfo{year}{2021}), \bibinfo{pages}{8780--8794}.
\newblock


\bibitem[\protect\citeauthoryear{Di~Fede, Rocchesso, Dow, and Andolina}{Di~Fede et~al\mbox{.}}{2022}]%
        {di2022idea}
\bibfield{author}{\bibinfo{person}{Giulia Di~Fede}, \bibinfo{person}{Davide Rocchesso}, \bibinfo{person}{Steven~P Dow}, {and} \bibinfo{person}{Salvatore Andolina}.} \bibinfo{year}{2022}\natexlab{}.
\newblock \showarticletitle{The idea machine: LLM-based expansion, rewriting, combination, and suggestion of ideas}. In \bibinfo{booktitle}{\emph{Proceedings of the 14th Conference on Creativity and Cognition}}. \bibinfo{pages}{623--627}.
\newblock


\bibitem[\protect\citeauthoryear{Dwairy, Dowell, and Stahl}{Dwairy et~al\mbox{.}}{2011}]%
        {dwairy2011application}
\bibfield{author}{\bibinfo{person}{Mai Dwairy}, \bibinfo{person}{Anthony~C Dowell}, {and} \bibinfo{person}{Jean-Claude Stahl}.} \bibinfo{year}{2011}\natexlab{}.
\newblock \showarticletitle{The application of foraging theory to the information searching behaviour of general practitioners}.
\newblock \bibinfo{journal}{\emph{BMC family practice}}  \bibinfo{volume}{12} (\bibinfo{year}{2011}), \bibinfo{pages}{1--8}.
\newblock


\bibitem[\protect\citeauthoryear{Feng, Li, and Zhang}{Feng et~al\mbox{.}}{2023a}]%
        {feng_understanding_2023}
\bibfield{author}{\bibinfo{person}{K.~J.~Kevin Feng}, \bibinfo{person}{Tony~W Li}, {and} \bibinfo{person}{Amy~X. Zhang}.} \bibinfo{year}{2023}\natexlab{a}.
\newblock \showarticletitle{Understanding {Collaborative} {Practices} and {Tools} of {Professional} {UX} {Practitioners} in {Software} {Organizations}}. In \bibinfo{booktitle}{\emph{Proceedings of the 2023 {CHI} {Conference} on {Human} {Factors} in {Computing} {Systems}}} \emph{(\bibinfo{series}{{CHI} '23})}. \bibinfo{publisher}{Association for Computing Machinery}, \bibinfo{address}{New York, NY, USA}, \bibinfo{pages}{1--20}.
\newblock
\showISBNx{978-1-4503-9421-5}
\urldef\tempurl%
\url{https://doi.org/10.1145/3544548.3581273}
\showDOI{\tempurl}


\bibitem[\protect\citeauthoryear{Feng and Zhang}{Feng and Zhang}{2022}]%
        {feng_handoffs_2022}
\bibfield{author}{\bibinfo{person}{K~J~Kevin Feng} {and} \bibinfo{person}{Amy~X Zhang}.} \bibinfo{year}{2022}\natexlab{}.
\newblock \showarticletitle{From {Handoffs} to {Co}-{Creation}: {Deepening} {Collaboration} between {Designers}, {Developers}, and {Data} {Science} {Workers} in {UX} {Design}}.
\newblock  (\bibinfo{year}{2022}).
\newblock


\bibitem[\protect\citeauthoryear{Feng, Zhu, Fu, Jampani, Akula, He, Basu, Wang, and Wang}{Feng et~al\mbox{.}}{2023b}]%
        {feng_layoutgpt_2023}
\bibfield{author}{\bibinfo{person}{Weixi Feng}, \bibinfo{person}{Wanrong Zhu}, \bibinfo{person}{Tsu-jui Fu}, \bibinfo{person}{Varun Jampani}, \bibinfo{person}{Arjun Akula}, \bibinfo{person}{Xuehai He}, \bibinfo{person}{Sugato Basu}, \bibinfo{person}{Xin~Eric Wang}, {and} \bibinfo{person}{William~Yang Wang}.} \bibinfo{year}{2023}\natexlab{b}.
\newblock \bibinfo{title}{{LayoutGPT}: {Compositional} {Visual} {Planning} and {Generation} with {Large} {Language} {Models}}.
\newblock
\newblock
\urldef\tempurl%
\url{https://doi.org/10.48550/arXiv.2305.15393}
\showDOI{\tempurl}
\newblock
\shownote{arXiv:2305.15393 [cs]}.


\bibitem[\protect\citeauthoryear{Feng, Fang, Cai, and Zhang}{Feng et~al\mbox{.}}{2021}]%
        {feng_guis2code_2021}
\bibfield{author}{\bibinfo{person}{Zhen Feng}, \bibinfo{person}{Jiaqi Fang}, \bibinfo{person}{Bo Cai}, {and} \bibinfo{person}{Yingtao Zhang}.} \bibinfo{year}{2021}\natexlab{}.
\newblock \showarticletitle{{GUIS2Code}: {A} {Computer} {Vision} {Tool} to {Generate} {Code} {Automatically} from {Graphical} {User} {Interface} {Sketches}}. In \bibinfo{booktitle}{\emph{Artificial {Neural} {Networks} and {Machine} {Learning} – {ICANN} 2021}} \emph{(\bibinfo{series}{Lecture {Notes} in {Computer} {Science}})}, \bibfield{editor}{\bibinfo{person}{Igor Farkaš}, \bibinfo{person}{Paolo Masulli}, \bibinfo{person}{Sebastian Otte}, {and} \bibinfo{person}{Stefan Wermter}} (Eds.). \bibinfo{publisher}{Springer International Publishing}, \bibinfo{address}{Cham}, \bibinfo{pages}{53--65}.
\newblock
\showISBNx{978-3-030-86365-4}
\urldef\tempurl%
\url{https://doi.org/10.1007/978-3-030-86365-4_5}
\showDOI{\tempurl}


\bibitem[\protect\citeauthoryear{Foundation}{Foundation}{[n.\,d.]}]%
        {idf_user_flow}
\bibfield{author}{\bibinfo{person}{Interaction~Design Foundation}.} \bibinfo{year}{[n.\,d.]}\natexlab{}.
\newblock \bibinfo{title}{What Are User Flows?}
\newblock
\newblock
\urldef\tempurl%
\url{https://www.interaction-design.org/literature/topics/user-flows}
\showURL{%
\tempurl}


\bibitem[\protect\citeauthoryear{Frost}{Frost}{2016}]%
        {frost_atomic_2016}
\bibfield{author}{\bibinfo{person}{Brad Frost}.} \bibinfo{year}{2016}\natexlab{}.
\newblock \bibinfo{booktitle}{\emph{Atomic {Design}}}.
\newblock
\urldef\tempurl%
\url{http://atomicdesign.bradfrost.com/}
\showURL{%
\tempurl}


\bibitem[\protect\citeauthoryear{Gassmann and Zeschky}{Gassmann and Zeschky}{2008}]%
        {gassmann2008opening}
\bibfield{author}{\bibinfo{person}{Oliver Gassmann} {and} \bibinfo{person}{Marco Zeschky}.} \bibinfo{year}{2008}\natexlab{}.
\newblock \showarticletitle{Opening up the solution space: The role of analogical thinking for breakthrough product innovation}.
\newblock \bibinfo{journal}{\emph{Creativity and Innovation Management}} \bibinfo{volume}{17}, \bibinfo{number}{2} (\bibinfo{year}{2008}), \bibinfo{pages}{97--106}.
\newblock


\bibitem[\protect\citeauthoryear{Gentner and Markman}{Gentner and Markman}{1997}]%
        {gentner1997structure}
\bibfield{author}{\bibinfo{person}{Dedre Gentner} {and} \bibinfo{person}{Arthur~B Markman}.} \bibinfo{year}{1997}\natexlab{}.
\newblock \showarticletitle{Structure mapping in analogy and similarity.}
\newblock \bibinfo{journal}{\emph{American psychologist}} \bibinfo{volume}{52}, \bibinfo{number}{1} (\bibinfo{year}{1997}), \bibinfo{pages}{45}.
\newblock


\bibitem[\protect\citeauthoryear{Girotra, Meincke, Terwiesch, and Ulrich}{Girotra et~al\mbox{.}}{2023}]%
        {girotra2023ideas}
\bibfield{author}{\bibinfo{person}{Karan Girotra}, \bibinfo{person}{Lennart Meincke}, \bibinfo{person}{Christian Terwiesch}, {and} \bibinfo{person}{Karl~T Ulrich}.} \bibinfo{year}{2023}\natexlab{}.
\newblock \showarticletitle{Ideas are dimes a dozen: Large language models for idea generation in innovation}.
\newblock \bibinfo{journal}{\emph{Available at SSRN 4526071}} (\bibinfo{year}{2023}).
\newblock


\bibitem[\protect\citeauthoryear{Goel, Vattam, Wiltgen, and Helms}{Goel et~al\mbox{.}}{2012}]%
        {goel2012cognitive}
\bibfield{author}{\bibinfo{person}{Ashok~K Goel}, \bibinfo{person}{Swaroop Vattam}, \bibinfo{person}{Bryan Wiltgen}, {and} \bibinfo{person}{Michael Helms}.} \bibinfo{year}{2012}\natexlab{}.
\newblock \showarticletitle{Cognitive, collaborative, conceptual and creative—four characteristics of the next generation of knowledge-based CAD systems: a study in biologically inspired design}.
\newblock \bibinfo{journal}{\emph{Computer-Aided Design}} \bibinfo{volume}{44}, \bibinfo{number}{10} (\bibinfo{year}{2012}), \bibinfo{pages}{879--900}.
\newblock


\bibitem[\protect\citeauthoryear{Gonzalez, Moran, Houde, He, Ross, Muller, Kunde, and Weisz}{Gonzalez et~al\mbox{.}}{2024}]%
        {gonzalez2024collaborative}
\bibfield{author}{\bibinfo{person}{Gabriel~Enrique Gonzalez}, \bibinfo{person}{Dario Andres~Silva Moran}, \bibinfo{person}{Stephanie Houde}, \bibinfo{person}{Jessica He}, \bibinfo{person}{Steven Ross}, \bibinfo{person}{Michael Muller}, \bibinfo{person}{Siya Kunde}, {and} \bibinfo{person}{Justin Weisz}.} \bibinfo{year}{2024}\natexlab{}.
\newblock \showarticletitle{Collaborative Canvas: A Tool for Exploring LLM Use in Group Ideation Tasks}. In \bibinfo{booktitle}{\emph{ACM International Conference on Intelligent User Interfaces}}.
\newblock


\bibitem[\protect\citeauthoryear{Gray}{Gray}{2016}]%
        {gray_its_2016}
\bibfield{author}{\bibinfo{person}{Colin~M. Gray}.} \bibinfo{year}{2016}\natexlab{}.
\newblock \showarticletitle{"{It}'s {More} of a {Mindset} {Than} a {Method}": {UX} {Practitioners}' {Conception} of {Design} {Methods}}. In \bibinfo{booktitle}{\emph{Proceedings of the 2016 {CHI} {Conference} on {Human} {Factors} in {Computing} {Systems}}} \emph{(\bibinfo{series}{{CHI} '16})}. \bibinfo{publisher}{Association for Computing Machinery}, \bibinfo{address}{New York, NY, USA}, \bibinfo{pages}{4044--4055}.
\newblock
\showISBNx{978-1-4503-3362-7}
\urldef\tempurl%
\url{https://doi.org/10.1145/2858036.2858410}
\showDOI{\tempurl}


\bibitem[\protect\citeauthoryear{Hassani, Silva, Unger, TajMazinani, and Mac~Feely}{Hassani et~al\mbox{.}}{2020}]%
        {hassani2020artificial}
\bibfield{author}{\bibinfo{person}{Hossein Hassani}, \bibinfo{person}{Emmanuel~Sirimal Silva}, \bibinfo{person}{Stephane Unger}, \bibinfo{person}{Maedeh TajMazinani}, {and} \bibinfo{person}{Stephen Mac~Feely}.} \bibinfo{year}{2020}\natexlab{}.
\newblock \showarticletitle{Artificial intelligence (AI) or intelligence augmentation (IA): what is the future?}
\newblock \bibinfo{journal}{\emph{Ai}} \bibinfo{volume}{1}, \bibinfo{number}{2} (\bibinfo{year}{2020}), \bibinfo{pages}{8}.
\newblock


\bibitem[\protect\citeauthoryear{Ho, Jain, and Abbeel}{Ho et~al\mbox{.}}{2020}]%
        {ho2020denoising}
\bibfield{author}{\bibinfo{person}{Jonathan Ho}, \bibinfo{person}{Ajay Jain}, {and} \bibinfo{person}{Pieter Abbeel}.} \bibinfo{year}{2020}\natexlab{}.
\newblock \showarticletitle{Denoising diffusion probabilistic models}.
\newblock \bibinfo{journal}{\emph{Advances in neural information processing systems}}  \bibinfo{volume}{33} (\bibinfo{year}{2020}), \bibinfo{pages}{6840--6851}.
\newblock


\bibitem[\protect\citeauthoryear{Huang, Canny, and Nichols}{Huang et~al\mbox{.}}{2019}]%
        {huang_swire_2019}
\bibfield{author}{\bibinfo{person}{Forrest Huang}, \bibinfo{person}{John~F. Canny}, {and} \bibinfo{person}{Jeffrey Nichols}.} \bibinfo{year}{2019}\natexlab{}.
\newblock \showarticletitle{Swire: {Sketch}-based {User} {Interface} {Retrieval}}. In \bibinfo{booktitle}{\emph{Proceedings of the 2019 {CHI} {Conference} on {Human} {Factors} in {Computing} {Systems}}} \emph{(\bibinfo{series}{{CHI} '19})}. \bibinfo{publisher}{Association for Computing Machinery}, \bibinfo{address}{New York, NY, USA}, \bibinfo{pages}{1--10}.
\newblock
\showISBNx{978-1-4503-5970-2}
\urldef\tempurl%
\url{https://doi.org/10.1145/3290605.3300334}
\showDOI{\tempurl}


\bibitem[\protect\citeauthoryear{Huang, Li, Zhou, Canny, and Li}{Huang et~al\mbox{.}}{2021}]%
        {huang_creating_2021}
\bibfield{author}{\bibinfo{person}{Forrest Huang}, \bibinfo{person}{Gang Li}, \bibinfo{person}{Xin Zhou}, \bibinfo{person}{John~F. Canny}, {and} \bibinfo{person}{Yang Li}.} \bibinfo{year}{2021}\natexlab{}.
\newblock \bibinfo{title}{Creating {User} {Interface} {Mock}-ups from {High}-{Level} {Text} {Descriptions} with {Deep}-{Learning} {Models}}.
\newblock
\newblock
\urldef\tempurl%
\url{https://doi.org/10.48550/arXiv.2110.07775}
\showDOI{\tempurl}
\newblock
\shownote{arXiv:2110.07775 [cs]}.


\bibitem[\protect\citeauthoryear{Kaplan}{Kaplan}{2023}]%
        {Kaplan_2023}
\bibfield{author}{\bibinfo{person}{Kate Kaplan}.} \bibinfo{year}{2023}\natexlab{}.
\newblock \bibinfo{title}{User Journeys vs. User Flows}.
\newblock
\newblock
\urldef\tempurl%
\url{https://www.nngroup.com/articles/user-journeys-vs-user-flows/}
\showURL{%
\tempurl}


\bibitem[\protect\citeauthoryear{Kim and Maher}{Kim and Maher}{2023}]%
        {kim_effect_2023}
\bibfield{author}{\bibinfo{person}{Jingoog Kim} {and} \bibinfo{person}{Mary~Lou Maher}.} \bibinfo{year}{2023}\natexlab{}.
\newblock \showarticletitle{The effect of {AI}-based inspiration on human design ideation}.
\newblock \bibinfo{journal}{\emph{International Journal of Design Creativity and Innovation}} \bibinfo{volume}{11}, \bibinfo{number}{2} (\bibinfo{date}{April} \bibinfo{year}{2023}), \bibinfo{pages}{81--98}.
\newblock
\showISSN{2165-0349}
\urldef\tempurl%
\url{https://doi.org/10.1080/21650349.2023.2167124}
\showDOI{\tempurl}
\newblock
\shownote{Publisher: Taylor \& Francis \_eprint: https://doi.org/10.1080/21650349.2023.2167124}.


\bibitem[\protect\citeauthoryear{Kim, Lee, Chang, and Kim}{Kim et~al\mbox{.}}{2023}]%
        {kim2023cells}
\bibfield{author}{\bibinfo{person}{Tae~Soo Kim}, \bibinfo{person}{Yoonjoo Lee}, \bibinfo{person}{Minsuk Chang}, {and} \bibinfo{person}{Juho Kim}.} \bibinfo{year}{2023}\natexlab{}.
\newblock \showarticletitle{Cells, generators, and lenses: Design framework for object-oriented interaction with large language models}. In \bibinfo{booktitle}{\emph{Proceedings of the 36th Annual ACM Symposium on User Interface Software and Technology}}. \bibinfo{pages}{1--18}.
\newblock


\bibitem[\protect\citeauthoryear{Kirillov, Mintun, Ravi, Mao, Rolland, Gustafson, Xiao, Whitehead, Berg, Lo, et~al\mbox{.}}{Kirillov et~al\mbox{.}}{2023}]%
        {kirillov2023segment}
\bibfield{author}{\bibinfo{person}{Alexander Kirillov}, \bibinfo{person}{Eric Mintun}, \bibinfo{person}{Nikhila Ravi}, \bibinfo{person}{Hanzi Mao}, \bibinfo{person}{Chloe Rolland}, \bibinfo{person}{Laura Gustafson}, \bibinfo{person}{Tete Xiao}, \bibinfo{person}{Spencer Whitehead}, \bibinfo{person}{Alexander~C Berg}, \bibinfo{person}{Wan-Yen Lo}, {et~al\mbox{.}}} \bibinfo{year}{2023}\natexlab{}.
\newblock \showarticletitle{Segment anything}. In \bibinfo{booktitle}{\emph{Proceedings of the IEEE/CVF International Conference on Computer Vision}}. \bibinfo{pages}{4015--4026}.
\newblock


\bibitem[\protect\citeauthoryear{Ko, Park, Jeon, Jo, Kim, and Seo}{Ko et~al\mbox{.}}{2023}]%
        {ko2023large}
\bibfield{author}{\bibinfo{person}{Hyung-Kwon Ko}, \bibinfo{person}{Gwanmo Park}, \bibinfo{person}{Hyeon Jeon}, \bibinfo{person}{Jaemin Jo}, \bibinfo{person}{Juho Kim}, {and} \bibinfo{person}{Jinwook Seo}.} \bibinfo{year}{2023}\natexlab{}.
\newblock \showarticletitle{Large-scale text-to-image generation models for visual artists’ creative works}. In \bibinfo{booktitle}{\emph{Proceedings of the 28th international conference on intelligent user interfaces}}. \bibinfo{pages}{919--933}.
\newblock


\bibitem[\protect\citeauthoryear{Koch, Lucero, Hegemann, and Oulasvirta}{Koch et~al\mbox{.}}{2019}]%
        {koch_may_2019}
\bibfield{author}{\bibinfo{person}{Janin Koch}, \bibinfo{person}{Andrés Lucero}, \bibinfo{person}{Lena Hegemann}, {and} \bibinfo{person}{Antti Oulasvirta}.} \bibinfo{year}{2019}\natexlab{}.
\newblock \showarticletitle{May {AI}? {Design} {Ideation} with {Cooperative} {Contextual} {Bandits}}. In \bibinfo{booktitle}{\emph{Proceedings of the 2019 {CHI} {Conference} on {Human} {Factors} in {Computing} {Systems}}} \emph{(\bibinfo{series}{{CHI} '19})}. \bibinfo{publisher}{Association for Computing Machinery}, \bibinfo{address}{New York, NY, USA}, \bibinfo{pages}{1--12}.
\newblock
\showISBNx{978-1-4503-5970-2}
\urldef\tempurl%
\url{https://doi.org/10.1145/3290605.3300863}
\showDOI{\tempurl}


\bibitem[\protect\citeauthoryear{Lamac}{Lamac}{2023}]%
        {lamac2023text}
\bibfield{author}{\bibinfo{person}{Radovan Lamac}.} \bibinfo{year}{2023}\natexlab{}.
\newblock \showarticletitle{Text-to-image AI as a tool for the designer’s ideation process}.
\newblock  (\bibinfo{year}{2023}).
\newblock


\bibitem[\protect\citeauthoryear{Lawrance, Bogart, Burnett, Bellamy, Rector, and Fleming}{Lawrance et~al\mbox{.}}{2010}]%
        {lawrance2010programmers}
\bibfield{author}{\bibinfo{person}{Joseph Lawrance}, \bibinfo{person}{Christopher Bogart}, \bibinfo{person}{Margaret Burnett}, \bibinfo{person}{Rachel Bellamy}, \bibinfo{person}{Kyle Rector}, {and} \bibinfo{person}{Scott~D Fleming}.} \bibinfo{year}{2010}\natexlab{}.
\newblock \showarticletitle{How programmers debug, revisited: An information foraging theory perspective}.
\newblock \bibinfo{journal}{\emph{IEEE Transactions on Software Engineering}} \bibinfo{volume}{39}, \bibinfo{number}{2} (\bibinfo{year}{2010}), \bibinfo{pages}{197--215}.
\newblock


\bibitem[\protect\citeauthoryear{Lewis, Perez, Piktus, Petroni, Karpukhin, Goyal, K{\"u}ttler, Lewis, Yih, Rockt{\"a}schel, et~al\mbox{.}}{Lewis et~al\mbox{.}}{2020}]%
        {lewis2020retrieval}
\bibfield{author}{\bibinfo{person}{Patrick Lewis}, \bibinfo{person}{Ethan Perez}, \bibinfo{person}{Aleksandra Piktus}, \bibinfo{person}{Fabio Petroni}, \bibinfo{person}{Vladimir Karpukhin}, \bibinfo{person}{Naman Goyal}, \bibinfo{person}{Heinrich K{\"u}ttler}, \bibinfo{person}{Mike Lewis}, \bibinfo{person}{Wen-tau Yih}, \bibinfo{person}{Tim Rockt{\"a}schel}, {et~al\mbox{.}}} \bibinfo{year}{2020}\natexlab{}.
\newblock \showarticletitle{Retrieval-augmented generation for knowledge-intensive nlp tasks}.
\newblock \bibinfo{journal}{\emph{Advances in Neural Information Processing Systems}}  \bibinfo{volume}{33} (\bibinfo{year}{2020}), \bibinfo{pages}{9459--9474}.
\newblock


\bibitem[\protect\citeauthoryear{Li and Li}{Li and Li}{2023}]%
        {li_spotlight_2023}
\bibfield{author}{\bibinfo{person}{Gang Li} {and} \bibinfo{person}{Yang Li}.} \bibinfo{year}{2023}\natexlab{}.
\newblock \bibinfo{title}{Spotlight: {Mobile} {UI} {Understanding} using {Vision}-{Language} {Models} with a {Focus}}.
\newblock
\newblock
\urldef\tempurl%
\url{https://doi.org/10.48550/arXiv.2209.14927}
\showDOI{\tempurl}
\newblock
\shownote{arXiv:2209.14927 [cs]}.


\bibitem[\protect\citeauthoryear{Li, Popowski, Mitchell, and Myers}{Li et~al\mbox{.}}{2021}]%
        {li_screen2vec_2021}
\bibfield{author}{\bibinfo{person}{Toby Jia-Jun Li}, \bibinfo{person}{Lindsay Popowski}, \bibinfo{person}{Tom Mitchell}, {and} \bibinfo{person}{Brad~A Myers}.} \bibinfo{year}{2021}\natexlab{}.
\newblock \showarticletitle{{Screen2Vec}: {Semantic} {Embedding} of {GUI} {Screens} and {GUI} {Components}}. In \bibinfo{booktitle}{\emph{Proceedings of the 2021 {CHI} {Conference} on {Human} {Factors} in {Computing} {Systems}}}. \bibinfo{publisher}{ACM}, \bibinfo{address}{Yokohama Japan}, \bibinfo{pages}{1--15}.
\newblock
\showISBNx{978-1-4503-8096-6}
\urldef\tempurl%
\url{https://doi.org/10.1145/3411764.3445049}
\showDOI{\tempurl}


\bibitem[\protect\citeauthoryear{Li, Li, He, Zheng, Li, and Guan}{Li et~al\mbox{.}}{2020}]%
        {li_widget_2020}
\bibfield{author}{\bibinfo{person}{Yang Li}, \bibinfo{person}{Gang Li}, \bibinfo{person}{Luheng He}, \bibinfo{person}{Jingjie Zheng}, \bibinfo{person}{Hong Li}, {and} \bibinfo{person}{Zhiwei Guan}.} \bibinfo{year}{2020}\natexlab{}.
\newblock \bibinfo{title}{Widget {Captioning}: {Generating} {Natural} {Language} {Description} for {Mobile} {User} {Interface} {Elements}}.
\newblock
\newblock
\urldef\tempurl%
\url{http://arxiv.org/abs/2010.04295}
\showURL{%
\tempurl}
\newblock
\shownote{arXiv:2010.04295 [cs]}.


\bibitem[\protect\citeauthoryear{Lu, Tong, Zhao, Zhang, and Li}{Lu et~al\mbox{.}}{2023}]%
        {lu2023ui}
\bibfield{author}{\bibinfo{person}{Yuwen Lu}, \bibinfo{person}{Ziang Tong}, \bibinfo{person}{Qinyi Zhao}, \bibinfo{person}{Chengzhi Zhang}, {and} \bibinfo{person}{Toby Jia-Jun Li}.} \bibinfo{year}{2023}\natexlab{}.
\newblock \showarticletitle{UI Layout Generation with LLMs Guided by UI Grammar}.
\newblock \bibinfo{journal}{\emph{arXiv preprint arXiv:2310.15455}} (\bibinfo{year}{2023}).
\newblock


\bibitem[\protect\citeauthoryear{Lu, Yang, Zhao, Zhang, and Li}{Lu et~al\mbox{.}}{2024}]%
        {lu2024ai}
\bibfield{author}{\bibinfo{person}{Yuwen Lu}, \bibinfo{person}{Yuewen Yang}, \bibinfo{person}{Qinyi Zhao}, \bibinfo{person}{Chengzhi Zhang}, {and} \bibinfo{person}{Toby Jia-Jun Li}.} \bibinfo{year}{2024}\natexlab{}.
\newblock \showarticletitle{AI Assistance for UX: A Literature Review Through Human-Centered AI}.
\newblock \bibinfo{journal}{\emph{arXiv preprint arXiv:2402.06089}} (\bibinfo{year}{2024}).
\newblock


\bibitem[\protect\citeauthoryear{Lu, Zhang, Zhang, and Li}{Lu et~al\mbox{.}}{2022}]%
        {lu_bridging_2022}
\bibfield{author}{\bibinfo{person}{Yuwen Lu}, \bibinfo{person}{Chengzhi Zhang}, \bibinfo{person}{Iris Zhang}, {and} \bibinfo{person}{Toby Jia-Jun Li}.} \bibinfo{year}{2022}\natexlab{}.
\newblock \showarticletitle{Bridging the {Gap} {Between} {UX} {Practitioners}' {Work} {Practices} and {AI}-{Enabled} {Design} {Support} {Tools}}. In \bibinfo{booktitle}{\emph{Extended {Abstracts} of the 2022 {CHI} {Conference} on {Human} {Factors} in {Computing} {Systems}}} \emph{(\bibinfo{series}{{CHI} {EA} '22})}. \bibinfo{publisher}{Association for Computing Machinery}, \bibinfo{address}{New York, NY, USA}, \bibinfo{pages}{1--7}.
\newblock
\showISBNx{978-1-4503-9156-6}
\urldef\tempurl%
\url{https://doi.org/10.1145/3491101.3519809}
\showDOI{\tempurl}


\bibitem[\protect\citeauthoryear{McDonald, Edwards, and Zhao}{McDonald et~al\mbox{.}}{2012}]%
        {mcdonald2012exploring}
\bibfield{author}{\bibinfo{person}{Sharon McDonald}, \bibinfo{person}{Helen~M Edwards}, {and} \bibinfo{person}{Tingting Zhao}.} \bibinfo{year}{2012}\natexlab{}.
\newblock \showarticletitle{Exploring think-alouds in usability testing: An international survey}.
\newblock \bibinfo{journal}{\emph{IEEE Transactions on Professional Communication}} \bibinfo{volume}{55}, \bibinfo{number}{1} (\bibinfo{year}{2012}), \bibinfo{pages}{2--19}.
\newblock


\bibitem[\protect\citeauthoryear{Mozaffari, Zhang, Cheng, and Guo}{Mozaffari et~al\mbox{.}}{2022}]%
        {mozaffari_ganspiration_2022}
\bibfield{author}{\bibinfo{person}{Mohammad~Amin Mozaffari}, \bibinfo{person}{Xinyuan Zhang}, \bibinfo{person}{Jinghui Cheng}, {and} \bibinfo{person}{Jin L.~C. Guo}.} \bibinfo{year}{2022}\natexlab{}.
\newblock \showarticletitle{{GANSpiration}: {Balancing} {Targeted} and {Serendipitous} {Inspiration} in {User} {Interface} {Design} with {Style}-{Based} {Generative} {Adversarial} {Network}}. In \bibinfo{booktitle}{\emph{{CHI} {Conference} on {Human} {Factors} in {Computing} {Systems}}}. \bibinfo{pages}{1--15}.
\newblock
\urldef\tempurl%
\url{https://doi.org/10.1145/3491102.3517511}
\showDOI{\tempurl}
\newblock
\shownote{arXiv:2203.03827 [cs]}.


\bibitem[\protect\citeauthoryear{Neil}{Neil}{2014}]%
        {neil2014mobile}
\bibfield{author}{\bibinfo{person}{Theresa Neil}.} \bibinfo{year}{2014}\natexlab{}.
\newblock \bibinfo{booktitle}{\emph{Mobile design pattern gallery: UI patterns for smartphone apps}}.
\newblock \bibinfo{publisher}{" O'Reilly Media, Inc."}.
\newblock


\bibitem[\protect\citeauthoryear{Nguyen, Vu, Pham, and Nguyen}{Nguyen et~al\mbox{.}}{2018}]%
        {nguyen2018deep}
\bibfield{author}{\bibinfo{person}{Tam~The Nguyen}, \bibinfo{person}{Phong~Minh Vu}, \bibinfo{person}{Hung~Viet Pham}, {and} \bibinfo{person}{Tung~Thanh Nguyen}.} \bibinfo{year}{2018}\natexlab{}.
\newblock \showarticletitle{Deep learning UI design patterns of mobile apps}. In \bibinfo{booktitle}{\emph{Proceedings of the 40th International Conference on Software Engineering: New Ideas and Emerging Results}}. \bibinfo{pages}{65--68}.
\newblock


\bibitem[\protect\citeauthoryear{Norman}{Norman}{2013}]%
        {norman2013design}
\bibfield{author}{\bibinfo{person}{Don Norman}.} \bibinfo{year}{2013}\natexlab{}.
\newblock \bibinfo{booktitle}{\emph{The design of everyday things: Revised and expanded edition}}.
\newblock \bibinfo{publisher}{Basic books}.
\newblock


\bibitem[\protect\citeauthoryear{Norman and Nielsen}{Norman and Nielsen}{1998}]%
        {norman1998definition}
\bibfield{author}{\bibinfo{person}{Don Norman} {and} \bibinfo{person}{Jakob Nielsen}.} \bibinfo{year}{1998}\natexlab{}.
\newblock \showarticletitle{The Definition of User Experience}.
\newblock \bibinfo{journal}{\emph{Nielsen Norman Group}} (\bibinfo{date}{Aug} \bibinfo{year}{1998}).
\newblock
\urldef\tempurl%
\url{https://www.nngroup.com/articles/definition-user-experience/}
\showURL{%
\tempurl}


\bibitem[\protect\citeauthoryear{North and Shneiderman}{North and Shneiderman}{2000}]%
        {north2000snap}
\bibfield{author}{\bibinfo{person}{Chris North} {and} \bibinfo{person}{Ben Shneiderman}.} \bibinfo{year}{2000}\natexlab{}.
\newblock \showarticletitle{Snap-together visualization: a user interface for coordinating visualizations via relational schemata}. In \bibinfo{booktitle}{\emph{Proceedings of the working conference on Advanced visual interfaces}}. \bibinfo{pages}{128--135}.
\newblock


\bibitem[\protect\citeauthoryear{Oppenlaender}{Oppenlaender}{2022}]%
        {oppenlaender2022creativity}
\bibfield{author}{\bibinfo{person}{Jonas Oppenlaender}.} \bibinfo{year}{2022}\natexlab{}.
\newblock \showarticletitle{The creativity of text-to-image generation}. In \bibinfo{booktitle}{\emph{Proceedings of the 25th International Academic Mindtrek Conference}}. \bibinfo{pages}{192--202}.
\newblock


\bibitem[\protect\citeauthoryear{Ozkan and Dogan}{Ozkan and Dogan}{2013}]%
        {ozkan2013cognitive}
\bibfield{author}{\bibinfo{person}{Ozgu Ozkan} {and} \bibinfo{person}{Fehmi Dogan}.} \bibinfo{year}{2013}\natexlab{}.
\newblock \showarticletitle{Cognitive strategies of analogical reasoning in design: Differences between expert and novice designers}.
\newblock \bibinfo{journal}{\emph{Design Studies}} \bibinfo{volume}{34}, \bibinfo{number}{2} (\bibinfo{year}{2013}), \bibinfo{pages}{161--192}.
\newblock


\bibitem[\protect\citeauthoryear{Paananen, Oppenlaender, and Visuri}{Paananen et~al\mbox{.}}{2023}]%
        {paananen2023using}
\bibfield{author}{\bibinfo{person}{Ville Paananen}, \bibinfo{person}{Jonas Oppenlaender}, {and} \bibinfo{person}{Aku Visuri}.} \bibinfo{year}{2023}\natexlab{}.
\newblock \showarticletitle{Using text-to-image generation for architectural design ideation}.
\newblock \bibinfo{journal}{\emph{International Journal of Architectural Computing}} (\bibinfo{year}{2023}), \bibinfo{pages}{14780771231222783}.
\newblock


\bibitem[\protect\citeauthoryear{Pernice}{Pernice}{2018}]%
        {pernice_2018}
\bibfield{author}{\bibinfo{person}{Kara Pernice}.} \bibinfo{year}{2018}\natexlab{}.
\newblock \bibinfo{title}{Affinity Diagramming: Collaboratively Sort UX Findings \& Design Ideas}.
\newblock
\newblock
\urldef\tempurl%
\url{https://www.nngroup.com/articles/affinity-diagram/}
\showURL{%
\tempurl}


\bibitem[\protect\citeauthoryear{Pirolli and Card}{Pirolli and Card}{1999}]%
        {pirolli1999information}
\bibfield{author}{\bibinfo{person}{Peter Pirolli} {and} \bibinfo{person}{Stuart Card}.} \bibinfo{year}{1999}\natexlab{}.
\newblock \showarticletitle{Information foraging.}
\newblock \bibinfo{journal}{\emph{Psychological review}} \bibinfo{volume}{106}, \bibinfo{number}{4} (\bibinfo{year}{1999}), \bibinfo{pages}{643}.
\newblock


\bibitem[\protect\citeauthoryear{Qu, Xiang, and Song}{Qu et~al\mbox{.}}{2023}]%
        {qu2023sketchdreamer}
\bibfield{author}{\bibinfo{person}{Zhiyu Qu}, \bibinfo{person}{Tao Xiang}, {and} \bibinfo{person}{Yi-Zhe Song}.} \bibinfo{year}{2023}\natexlab{}.
\newblock \showarticletitle{SketchDreamer: Interactive Text-Augmented Creative Sketch Ideation}.
\newblock \bibinfo{journal}{\emph{arXiv preprint arXiv:2308.14191}} (\bibinfo{year}{2023}).
\newblock


\bibitem[\protect\citeauthoryear{Russell and Chi}{Russell and Chi}{2014}]%
        {russell2014looking}
\bibfield{author}{\bibinfo{person}{Daniel~M Russell} {and} \bibinfo{person}{Ed~H Chi}.} \bibinfo{year}{2014}\natexlab{}.
\newblock \showarticletitle{Looking back: Retrospective study methods for HCI}.
\newblock In \bibinfo{booktitle}{\emph{Ways of Knowing in HCI}}. \bibinfo{publisher}{Springer}, \bibinfo{pages}{373--393}.
\newblock


\bibitem[\protect\citeauthoryear{Sandstrom}{Sandstrom}{1994}]%
        {sandstrom1994optimal}
\bibfield{author}{\bibinfo{person}{Pamela~Effrein Sandstrom}.} \bibinfo{year}{1994}\natexlab{}.
\newblock \showarticletitle{An optimal foraging approach to information seeking and use}.
\newblock \bibinfo{journal}{\emph{The library quarterly}} \bibinfo{volume}{64}, \bibinfo{number}{4} (\bibinfo{year}{1994}), \bibinfo{pages}{414--449}.
\newblock


\bibitem[\protect\citeauthoryear{Schoop, Zhou, Li, Chen, Hartmann, and Li}{Schoop et~al\mbox{.}}{2022}]%
        {schoop_predicting_2022}
\bibfield{author}{\bibinfo{person}{Eldon Schoop}, \bibinfo{person}{Xin Zhou}, \bibinfo{person}{Gang Li}, \bibinfo{person}{Zhourong Chen}, \bibinfo{person}{Bjoern Hartmann}, {and} \bibinfo{person}{Yang Li}.} \bibinfo{year}{2022}\natexlab{}.
\newblock \showarticletitle{Predicting and {Explaining} {Mobile} {UI} {Tappability} with {Vision} {Modeling} and {Saliency} {Analysis}}. In \bibinfo{booktitle}{\emph{{CHI} {Conference} on {Human} {Factors} in {Computing} {Systems}}}. \bibinfo{publisher}{ACM}, \bibinfo{address}{New Orleans LA USA}, \bibinfo{pages}{1--21}.
\newblock
\showISBNx{978-1-4503-9157-3}
\urldef\tempurl%
\url{https://doi.org/10.1145/3491102.3517497}
\showDOI{\tempurl}


\bibitem[\protect\citeauthoryear{Shaer, Cooper, Mokryn, Kun, and Shoshan}{Shaer et~al\mbox{.}}{2024}]%
        {shaer2024ai}
\bibfield{author}{\bibinfo{person}{Orit Shaer}, \bibinfo{person}{Angelora Cooper}, \bibinfo{person}{Osnat Mokryn}, \bibinfo{person}{Andrew~L Kun}, {and} \bibinfo{person}{Hagit~Ben Shoshan}.} \bibinfo{year}{2024}\natexlab{}.
\newblock \showarticletitle{AI-Augmented Brainwriting: Investigating the use of LLMs in group ideation}.
\newblock \bibinfo{journal}{\emph{arXiv preprint arXiv:2402.14978}} (\bibinfo{year}{2024}).
\newblock


\bibitem[\protect\citeauthoryear{Shneiderman}{Shneiderman}{2022}]%
        {shneiderman_human-centered_2022}
\bibfield{author}{\bibinfo{person}{Ben Shneiderman}.} \bibinfo{year}{2022}\natexlab{}.
\newblock \bibinfo{booktitle}{\emph{Human-{Centered} {AI}}}.
\newblock \bibinfo{publisher}{Oxford University Press}.
\newblock
\showISBNx{978-0-19-266000-8}
\newblock
\shownote{Google-Books-ID: mSRXEAAAQBAJ}.


\bibitem[\protect\citeauthoryear{Singer}{Singer}{[n.\,d.]}]%
        {singer_breadboard}
\bibfield{author}{\bibinfo{person}{Ryan Singer}.} \bibinfo{year}{[n.\,d.]}\natexlab{}.
\newblock \bibinfo{title}{Find the Elements | Shape Up}.
\newblock
\newblock
\urldef\tempurl%
\url{https://basecamp.com/shapeup/1.3-chapter-04}
\showURL{%
\tempurl}


\bibitem[\protect\citeauthoryear{Suh, Chen, Min, Li, and Xia}{Suh et~al\mbox{.}}{2023}]%
        {suh2023structured}
\bibfield{author}{\bibinfo{person}{Sangho Suh}, \bibinfo{person}{Meng Chen}, \bibinfo{person}{Bryan Min}, \bibinfo{person}{Toby Jia-Jun Li}, {and} \bibinfo{person}{Haijun Xia}.} \bibinfo{year}{2023}\natexlab{}.
\newblock \showarticletitle{Structured Generation and Exploration of Design Space with Large Language Models for Human-AI Co-Creation}.
\newblock \bibinfo{journal}{\emph{arXiv preprint arXiv:2310.12953}} (\bibinfo{year}{2023}).
\newblock


\bibitem[\protect\citeauthoryear{Swearngin, Dontcheva, Li, Brandt, Dixon, and Ko}{Swearngin et~al\mbox{.}}{2018}]%
        {swearngin_rewire_2018}
\bibfield{author}{\bibinfo{person}{Amanda Swearngin}, \bibinfo{person}{Mira Dontcheva}, \bibinfo{person}{Wilmot Li}, \bibinfo{person}{Joel Brandt}, \bibinfo{person}{Morgan Dixon}, {and} \bibinfo{person}{Amy~J. Ko}.} \bibinfo{year}{2018}\natexlab{}.
\newblock \showarticletitle{Rewire: {Interface} {Design} {Assistance} from {Examples}}. In \bibinfo{booktitle}{\emph{Proceedings of the 2018 {CHI} {Conference} on {Human} {Factors} in {Computing} {Systems}}} \emph{(\bibinfo{series}{{CHI} '18})}. \bibinfo{publisher}{Association for Computing Machinery}, \bibinfo{pages}{1--12}.
\newblock
\showISBNx{978-1-4503-5620-6}
\urldef\tempurl%
\url{https://doi.org/10.1145/3173574.3174078}
\showDOI{\tempurl}
\newblock
\shownote{Place: New York, NY, USA}.


\bibitem[\protect\citeauthoryear{Swearngin, Wang, Oleson, Fogarty, and Ko}{Swearngin et~al\mbox{.}}{2020}]%
        {swearngin_scout_2020}
\bibfield{author}{\bibinfo{person}{Amanda Swearngin}, \bibinfo{person}{Chenglong Wang}, \bibinfo{person}{Alannah Oleson}, \bibinfo{person}{James Fogarty}, {and} \bibinfo{person}{Amy~J. Ko}.} \bibinfo{year}{2020}\natexlab{}.
\newblock \showarticletitle{Scout: {Rapid} {Exploration} of {Interface} {Layout} {Alternatives} through {High}-{Level} {Design} {Constraints}}. In \bibinfo{booktitle}{\emph{Proceedings of the 2020 {CHI} {Conference} on {Human} {Factors} in {Computing} {Systems}}}. \bibinfo{pages}{1--13}.
\newblock
\urldef\tempurl%
\url{https://doi.org/10.1145/3313831.3376593}
\showDOI{\tempurl}
\newblock
\shownote{arXiv:2001.05424 [cs]}.


\bibitem[\protect\citeauthoryear{Tankala and Joyce}{Tankala and Joyce}{[n.\,d.]}]%
        {Tankala_Joyce}
\bibfield{author}{\bibinfo{person}{Samiha Tankala} {and} \bibinfo{person}{Alita Joyce}.} \bibinfo{year}{[n.\,d.]}\natexlab{}.
\newblock \bibinfo{title}{Design-Pattern Guidelines: Study Guide}.
\newblock
\newblock
\urldef\tempurl%
\url{https://www.nngroup.com/articles/design-pattern-guidelines/}
\showURL{%
\tempurl}


\bibitem[\protect\citeauthoryear{Tidwell}{Tidwell}{2010}]%
        {tidwell2010designing}
\bibfield{author}{\bibinfo{person}{Jenifer Tidwell}.} \bibinfo{year}{2010}\natexlab{}.
\newblock \bibinfo{booktitle}{\emph{Designing interfaces: Patterns for effective interaction design}}.
\newblock \bibinfo{publisher}{" O'Reilly Media, Inc."}.
\newblock


\bibitem[\protect\citeauthoryear{Vattam and Goel}{Vattam and Goel}{2011}]%
        {vattam2011foraging}
\bibfield{author}{\bibinfo{person}{Swaroop~S Vattam} {and} \bibinfo{person}{Ashok~K Goel}.} \bibinfo{year}{2011}\natexlab{}.
\newblock \showarticletitle{Foraging for inspiration: understanding and supporting the online information seeking practices of biologically inspired designers}. In \bibinfo{booktitle}{\emph{International Design Engineering Technical Conferences and Computers and Information in Engineering Conference}}, Vol.~\bibinfo{volume}{54860}. \bibinfo{pages}{177--186}.
\newblock


\bibitem[\protect\citeauthoryear{Vigo and Harper}{Vigo and Harper}{2013}]%
        {vigo2013challenging}
\bibfield{author}{\bibinfo{person}{Markel Vigo} {and} \bibinfo{person}{Simon Harper}.} \bibinfo{year}{2013}\natexlab{}.
\newblock \showarticletitle{Challenging information foraging theory: Screen reader users are not always driven by information scent}. In \bibinfo{booktitle}{\emph{Proceedings of the 24th acm conference on hypertext and social media}}. \bibinfo{pages}{60--68}.
\newblock


\bibitem[\protect\citeauthoryear{Wang, Li, and Li}{Wang et~al\mbox{.}}{2023}]%
        {wang_enabling_2023}
\bibfield{author}{\bibinfo{person}{Bryan Wang}, \bibinfo{person}{Gang Li}, {and} \bibinfo{person}{Yang Li}.} \bibinfo{year}{2023}\natexlab{}.
\newblock \bibinfo{title}{Enabling {Conversational} {Interaction} with {Mobile} {UI} using {Large} {Language} {Models}}.
\newblock
\newblock
\urldef\tempurl%
\url{http://arxiv.org/abs/2209.08655}
\showURL{%
\tempurl}
\newblock
\shownote{arXiv:2209.08655 [cs]}.


\bibitem[\protect\citeauthoryear{Wang, Li, Zhou, Chen, Grossman, and Li}{Wang et~al\mbox{.}}{2021}]%
        {wang_screen2words_2021}
\bibfield{author}{\bibinfo{person}{Bryan Wang}, \bibinfo{person}{Gang Li}, \bibinfo{person}{Xin Zhou}, \bibinfo{person}{Zhourong Chen}, \bibinfo{person}{Tovi Grossman}, {and} \bibinfo{person}{Yang Li}.} \bibinfo{year}{2021}\natexlab{}.
\newblock \bibinfo{title}{{Screen2Words}: {Automatic} {Mobile} {UI} {Summarization} with {Multimodal} {Learning}}.
\newblock
\newblock
\urldef\tempurl%
\url{http://arxiv.org/abs/2108.03353}
\showURL{%
\tempurl}
\newblock
\shownote{arXiv:2108.03353 [cs]}.


\bibitem[\protect\citeauthoryear{Wei, Courbis, Lambolais, Xu, Bernard, and Dray}{Wei et~al\mbox{.}}{2023}]%
        {wei2023boosting}
\bibfield{author}{\bibinfo{person}{Jialiang Wei}, \bibinfo{person}{Anne-Lise Courbis}, \bibinfo{person}{Thomas Lambolais}, \bibinfo{person}{Binbin Xu}, \bibinfo{person}{Pierre~Louis Bernard}, {and} \bibinfo{person}{G{\'e}rard Dray}.} \bibinfo{year}{2023}\natexlab{}.
\newblock \showarticletitle{Boosting gui prototyping with diffusion models}. In \bibinfo{booktitle}{\emph{2023 IEEE 31st International Requirements Engineering Conference (RE)}}. IEEE, \bibinfo{pages}{275--280}.
\newblock


\bibitem[\protect\citeauthoryear{Wei, Wang, Schuurmans, Bosma, Xia, Chi, Le, Zhou, et~al\mbox{.}}{Wei et~al\mbox{.}}{2022}]%
        {wei2022chain}
\bibfield{author}{\bibinfo{person}{Jason Wei}, \bibinfo{person}{Xuezhi Wang}, \bibinfo{person}{Dale Schuurmans}, \bibinfo{person}{Maarten Bosma}, \bibinfo{person}{Fei Xia}, \bibinfo{person}{Ed Chi}, \bibinfo{person}{Quoc~V Le}, \bibinfo{person}{Denny Zhou}, {et~al\mbox{.}}} \bibinfo{year}{2022}\natexlab{}.
\newblock \showarticletitle{Chain-of-thought prompting elicits reasoning in large language models}.
\newblock \bibinfo{journal}{\emph{Advances in neural information processing systems}}  \bibinfo{volume}{35} (\bibinfo{year}{2022}), \bibinfo{pages}{24824--24837}.
\newblock


\bibitem[\protect\citeauthoryear{Wu, Wang, Shen, Peng, Nichols, and Bigham}{Wu et~al\mbox{.}}{2023}]%
        {wu_webui_2023}
\bibfield{author}{\bibinfo{person}{Jason Wu}, \bibinfo{person}{Siyan Wang}, \bibinfo{person}{Siman Shen}, \bibinfo{person}{Yi-Hao Peng}, \bibinfo{person}{Jeffrey Nichols}, {and} \bibinfo{person}{Jeffrey~P. Bigham}.} \bibinfo{year}{2023}\natexlab{}.
\newblock \bibinfo{title}{{WebUI}: {A} {Dataset} for {Enhancing} {Visual} {UI} {Understanding} with {Web} {Semantics}}.
\newblock
\newblock
\urldef\tempurl%
\url{https://doi.org/10.48550/arXiv.2301.13280}
\showDOI{\tempurl}
\newblock
\shownote{arXiv:2301.13280 [cs]}.


\bibitem[\protect\citeauthoryear{Wu, Terry, and Cai}{Wu et~al\mbox{.}}{2022}]%
        {wu2022ai}
\bibfield{author}{\bibinfo{person}{Tongshuang Wu}, \bibinfo{person}{Michael Terry}, {and} \bibinfo{person}{Carrie~Jun Cai}.} \bibinfo{year}{2022}\natexlab{}.
\newblock \showarticletitle{Ai chains: Transparent and controllable human-ai interaction by chaining large language model prompts}. In \bibinfo{booktitle}{\emph{Proceedings of the 2022 CHI conference on human factors in computing systems}}. \bibinfo{pages}{1--22}.
\newblock


\bibitem[\protect\citeauthoryear{Yan, Yang, Zhu, Lin, Li, Wang, Yang, Zhong, McAuley, Gao, et~al\mbox{.}}{Yan et~al\mbox{.}}{2023}]%
        {yan2023gpt}
\bibfield{author}{\bibinfo{person}{An Yan}, \bibinfo{person}{Zhengyuan Yang}, \bibinfo{person}{Wanrong Zhu}, \bibinfo{person}{Kevin Lin}, \bibinfo{person}{Linjie Li}, \bibinfo{person}{Jianfeng Wang}, \bibinfo{person}{Jianwei Yang}, \bibinfo{person}{Yiwu Zhong}, \bibinfo{person}{Julian McAuley}, \bibinfo{person}{Jianfeng Gao}, {et~al\mbox{.}}} \bibinfo{year}{2023}\natexlab{}.
\newblock \showarticletitle{Gpt-4v in wonderland: Large multimodal models for zero-shot smartphone gui navigation}.
\newblock \bibinfo{journal}{\emph{arXiv preprint arXiv:2311.07562}} (\bibinfo{year}{2023}).
\newblock


\bibitem[\protect\citeauthoryear{Yang, Zhang, Li, Zou, Li, and Gao}{Yang et~al\mbox{.}}{2023}]%
        {Yang_Zhang_Li_Zou_Li_Gao_2023}
\bibfield{author}{\bibinfo{person}{Jianwei Yang}, \bibinfo{person}{Hao Zhang}, \bibinfo{person}{Feng Li}, \bibinfo{person}{Xueyan Zou}, \bibinfo{person}{Chunyuan Li}, {and} \bibinfo{person}{Jianfeng Gao}.} \bibinfo{year}{2023}\natexlab{}.
\newblock \showarticletitle{Set-of-Mark Prompting Unleashes Extraordinary Visual Grounding in GPT-4V}.
\newblock  \bibinfo{number}{arXiv:2310.11441} (\bibinfo{date}{Nov.} \bibinfo{year}{2023}).
\newblock
\urldef\tempurl%
\url{https://doi.org/10.48550/arXiv.2310.11441}
\showDOI{\tempurl}
\newblock
\shownote{arXiv:2310.11441 [cs]}.


\bibitem[\protect\citeauthoryear{YIN, GU, MAR, ZHANG, and DOW}{YIN et~al\mbox{.}}{2024}]%
        {yin2024jamplate}
\bibfield{author}{\bibinfo{person}{JIAYU YIN}, \bibinfo{person}{CATHERINE GU}, \bibinfo{person}{JENNY MAR}, \bibinfo{person}{SYDNEY ZHANG}, {and} \bibinfo{person}{STEVEN~P DOW}.} \bibinfo{year}{2024}\natexlab{}.
\newblock \showarticletitle{Jamplate: Exploring LLM-Enhanced Templates for Idea Reflection}.
\newblock  (\bibinfo{year}{2024}).
\newblock


\bibitem[\protect\citeauthoryear{Zhao, Chen, Liu, and Zhu}{Zhao et~al\mbox{.}}{2021}]%
        {zhao_guigan_2021}
\bibfield{author}{\bibinfo{person}{Tianming Zhao}, \bibinfo{person}{Chunyang Chen}, \bibinfo{person}{Yuanning Liu}, {and} \bibinfo{person}{Xiaodong Zhu}.} \bibinfo{year}{2021}\natexlab{}.
\newblock \bibinfo{title}{{GUIGAN}: {Learning} to {Generate} {GUI} {Designs} {Using} {Generative} {Adversarial} {Networks}}.
\newblock
\newblock
\urldef\tempurl%
\url{https://doi.org/10.48550/arXiv.2101.09978}
\showDOI{\tempurl}
\newblock
\shownote{arXiv:2101.09978 [cs]}.


\bibitem[\protect\citeauthoryear{Zhong, Li, and Li}{Zhong et~al\mbox{.}}{2021}]%
        {zhong2021spacewalker}
\bibfield{author}{\bibinfo{person}{Mingyuan Zhong}, \bibinfo{person}{Gang Li}, {and} \bibinfo{person}{Yang Li}.} \bibinfo{year}{2021}\natexlab{}.
\newblock \showarticletitle{Spacewalker: Rapid UI design exploration using lightweight markup enhancement and crowd genetic programming}. In \bibinfo{booktitle}{\emph{Proceedings of the 2021 CHI Conference on Human Factors in Computing Systems}}. \bibinfo{pages}{1--11}.
\newblock


\end{thebibliography}
